\documentclass[twocolumn, showpacs,preprintnumbers,amsmath,amssymb,superscriptaddress,prc]{revtex4}
\usepackage{graphicx}% Include figure files
\graphicspath{{fig/}}
\usepackage{dcolumn}% Align table columns on decimal point
\usepackage{bm}% bold math
\usepackage[usenames]{color} % add some color
\definecolor{DarkBlue}{rgb}{0.1,0,0.55} 
\definecolor{DarkRed}{rgb}{.139,.055,.055}
\definecolor{DarkGreen}{rgb}{.0,.40,.0}
\usepackage{pxfonts}
\usepackage{pifont}

\newcommand{\leff}{{\small $\mathcal{L}_\textrm{eff}$}}
\newcommand{\lefftitle}{{ $\mathcal{L}_\textrm{EFF}$}}
\newcommand{\leffeq}{  \mbox{\fontsize{9}{11}\selectfont $\mathcal{L}_\textrm{eff}$} }
\newcommand{\kevr}{keV$_\textrm{r}$}

\begin{document}
% \switchlinenumbers    \linenumbers
\title{Scintillation efficiency and ionization yield of liquid xenon 
for mono-energetic nuclear recoils down to 4 keV}% Force line breaks with \\

\author{A.~Manzur}		
\author{A.~Curioni}
\altaffiliation[Current address: ]{Institute for Particle Physics, ETH Zurich, 8093 Zurich, Switzerland}
% CERN - European Organization for Nuclear Research}
\author{L.~Kastens}		
\author{D.N.~McKinsey} 	
\email[Corresponding author: ]{daniel.mckinsey@yale.edu}
\author{K.~Ni}	
\altaffiliation[Current address: ]{Department of Physics, Shanghai Jiao Tong University, Shanghai, China}
\author{T.~Wongjirad}	
\altaffiliation[Current address: ]{Department of Physics, Duke University, Durham, NC, USA}
\affiliation{Department of Physics, Yale 
University, P.O. Box 208120, New Haven, CT 06520, USA}

\date{\today}% It is always \today, today,

\begin{abstract} 
Liquid Xenon (LXe) is an excellent material for experiments designed
to detect dark matter in the form of Weakly Interacting Massive
Particles (WIMPs).  A low energy detection threshold is essential for a
sensitive WIMP search.  The understanding of the relative scintillation
efficiency (\leff) and ionization yield of low energy nuclear recoils
in LXe is limited for energies below 10 keV. In this paper,
we present new measurements that extend the energy down to 4 keV,
finding that \leff~ decreases with decreasing energy. We also measure the quenching of
scintillation efficiency due to the electric field in LXe, finding no
significant field dependence.
\end{abstract} 

% uncomment for revtex
% \keywords{Suggested keywords}%Use showkeys class option if keyword
\maketitle

\section{Introduction} %%%%%
Liquid xenon is increasingly used as the detection material in direct
searches for WIMP dark matter  \cite{Jungman:96}.  Recent developments in two-phase (gas/liquid)
xenon detectors \cite{Alner:07,Angle:08, Lebedenko:08} has resulted in stringent
limits on the WIMP-nucleon cross-section, constraining theories of 
physics beyond the standard model, such as supersymmetry.  
WIMPs will deposit a small amount of energy in the LXe through 
elastic scatters with  xenon nuclei. Part of the 
deposited energy is converted into observable signals of 
scintillation light and ionization electrons. The rest of the energy 
is converted into heat and can not be easily measured. Understanding 
these effects  will help  
determine nuclear recoil energies and ultimately play a part 
in determining the WIMP-nucleon cross-section.

\begin{figure}[htbp]
	\begin{center}
		\includegraphics[width=7cm]{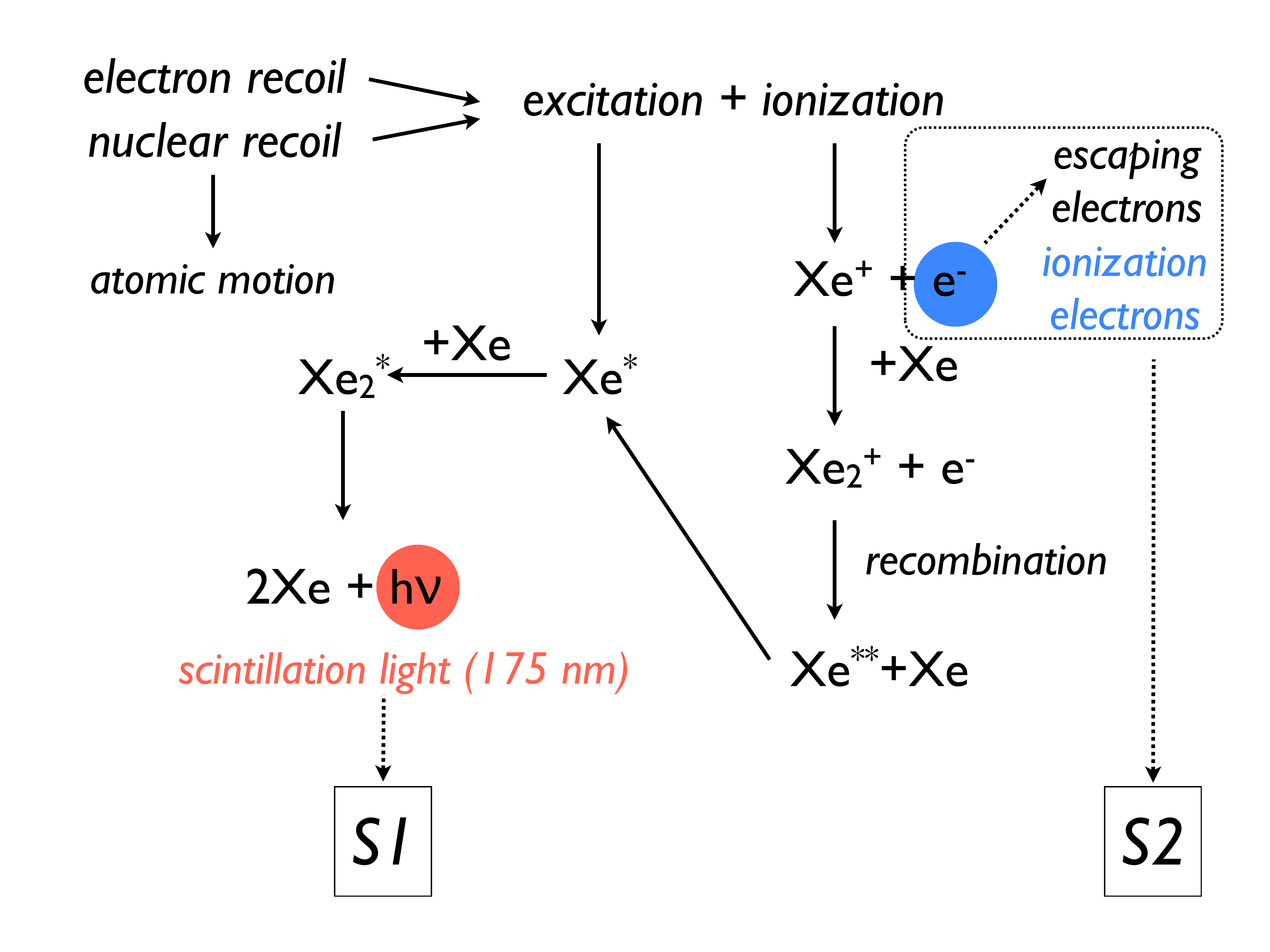}
		\caption{(Color online) Illustration of the signal production and collection in a two-phase xenon detector.}
		\label{fig:s1s2}
	\end{center}
\end{figure}

In a two-phase xenon detector, two signals are measured. The first is 
the direct scintillation light, denoted as $S1$. The second is the 
proportional scintillation light in the gas phase, denoted as $S2$, 
which is proportional to the ionization electrons that survive 
electron-ion recombination and are extracted into the gas. 
Figure~\ref{fig:s1s2} gives an illustration of the signal production 
and collection in a two-phase xenon detector.

For a given event in the LXe, the nuclear recoil energy can
be determined based on the scintillation signal $S1$
\cite{Alner:07,Angle:08}. However, it is much more convenient to calibrate the detector using 
 electron recoil events. The tradition in the field  \cite{Arneodo:00,Akimov:02,Aprile:05,Chepel:06,Aprile:08} is to base the energy calibration on 122~keV electron recoils from a $^{57}$Co gamma source.
 The \textit{relative scintillation
efficiency}, \leff, defined as the ratio between the electron equivalent energy ($E_{ee}$) and the true nuclear recoil energy ($E_r$), becomes  necessary for determining the nuclear energy scale and, therefore, the WIMP detection sensitivity.
$E_{ee}$ is  inferred from the scintillation
signal yield  due to  monoenergetic electron recoils. \leff~ has no
units and is defined at zero electric field in LXe relative to 122~keV gamma rays.

If an electric field is applied to the LXe, the 
scintillation yields for both electron and nuclear recoils are 
suppressed by additional factors $S_e$ and $S_n$, respectively. The 
relative scintillation efficiency can be calculated as
\begin{equation}
 \leffeq = E_{ee}/ E_{r} \cdot S_e/S_n \\
\end{equation}
The quantity $S_e$ for 122~keV electron recoils from a $^{57}$Co
source has been  measured very accurately \cite{Aprile:06}. 
$S_n$ has been measured for  56~keV nuclear recoils, with electric fields up to a few kV/cm
in LXe \cite{Aprile:06,Aprile:05}, but no measurement is available for
nuclear recoils at other energies.

% Since electron recoil energy calibrations are easier than nuclear
% recoil energy calibrations, knowledge of the scintillation efficiency
% can be used to determine the nuclear recoil energy scale, which is
 % important for calculating WIMP detection sensitivity.  The
% tradition in the field \cite{Arneodo:00,Akimov:02,Aprile:05,Chepel:06,Aprile:08} is to base the energy calibration for $E_{ee}$
% on 122~keV electron recoils .  
 % relative to 122~keV electron recoils. 
% The scintillation
% signal yield has the units of photoelectrons per keV. 

Two methods have been utilized to determine \leff~ as a function of
energy: a) Using a fixed-energy neutron beam experiment, detecting
neutrons that scatter in the LXe
at a known scattering angle, and b) Comparing
neutron calibration data to Monte Carlo simulations without tagging
the scattered neutron.  Using method a), \leff~ has 
been measured by a number of groups for nuclear recoils above 10~keV
\cite{Arneodo:00, Akimov:02, Aprile:05}.  There are also two
measurements of this type reporting results below 10~keV, one
suggesting an increasing \leff~ with decreasing energies
\cite{Chepel:06}, and another indicating a roughly constant \leff~ of $ 0.15$ down
to 5~keV \cite{Aprile:08}.  The XENON10 and the ZEPLIN-III
collaborations have also determined \leff~ with method b)
\cite{Sorensen:08, Sorensen:09, Lebedenko:08}.  In the XENON10 analysis, \leff~
does not decrease much at low energies, while in the ZEPLIN-III
measurement, the data imply a precipitous drop at low energies.

In this paper, we report on measurements of \leff~ at zero field and
$S_n$ at two different fields (0.73~kV/cm and 1.5~kV/cm) for nuclear
recoils between 4 and 66~keV in a single phase detector ($S1$-only). 
We repeated these measurements using a dual phase detector ($S1$ and
$S2$ signals) at 1~kV/cm in the liquid and 10~kV/cm in the gas as well
as 4~kV/cm in the liquid and 8~kV/cm in the gas.  With the dual phase
detector we also measured the ionization signal yield for nuclear
recoils.  We present the experimental apparatus in
Section~\ref{sec:exp}, the data analysis in
Section~\ref{sec:analysis}, the results in Section \ref{sec:results},
a theoretical model of \leff~in Section \ref{sec:model} and a discussion
of the results in Section~\ref{sec:discussion}.

\section{Experimental Apparatus}
\label{sec:exp}

\begin{figure}[htbp]
	\begin{center}
		\includegraphics[width=8cm]{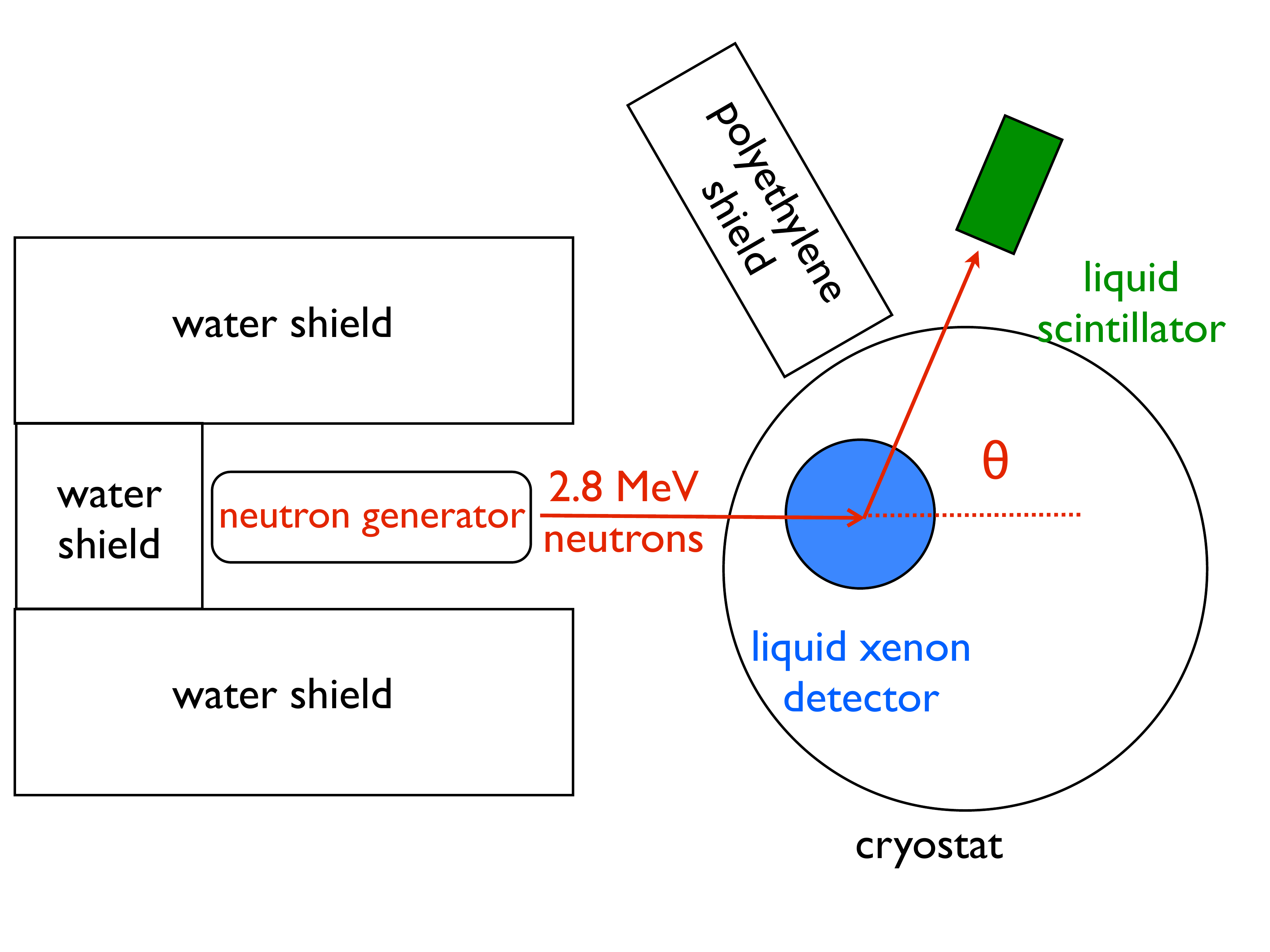}
		\caption{(Color online) The setup for nuclear recoil scintillation efficiency 
		measurement in LXe. Not drawn to scale.}
		\label{fig:setup}
	\end{center}
\end{figure}

\begin{figure}[htbp]
	\centering
		\includegraphics[width=8.0cm]{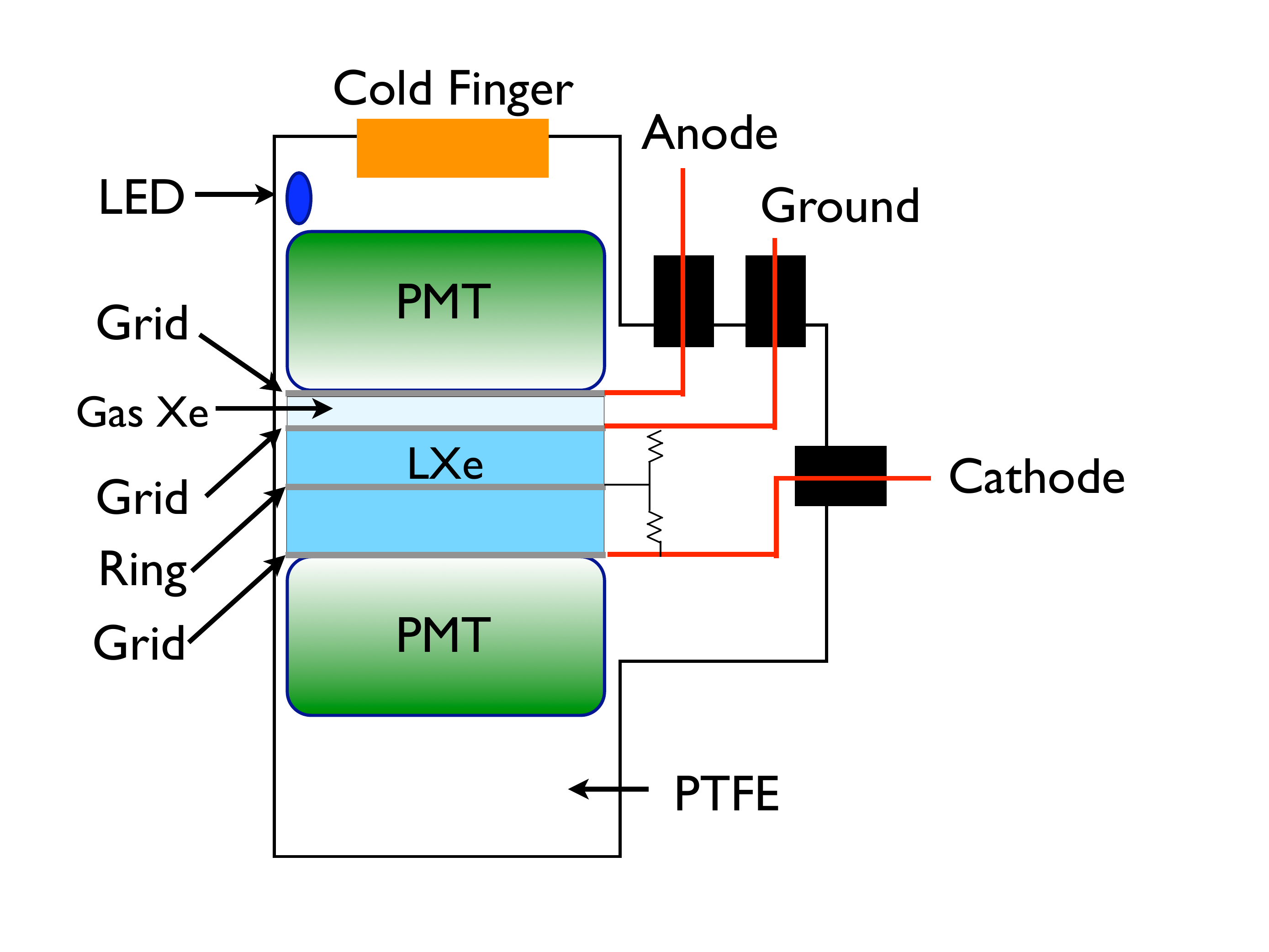}
	\caption{(Color online) Schematic of the dual phase LXe detector. Spaces 
	not drawn were filled with PTFE pieces. PMTs, LXe and xenon gas 
	regions drawn to scale.}
	\label{fig:geant4_cell}
\end{figure}

The measurement was performed with a setup comprising  a 
deuterium-deuterium neutron generator \cite{Chichester:07}, a LXe
 detector, and an organic liquid scintillator detector, as shown 
in Figure~\ref{fig:setup}. The neutron generator produces 2.8~MeV 
neutrons at a rate of $10^6$ n/s. The liquid scintillator detector is 
a BC501A organic scintillator module 3.8~cm in diameter and 3.8~cm in 
height, viewed by a photomultiplier (PMT). Both the neutron generator 
and the organic scintillator detector have previously been used to perform 
measurements of nuclear recoils in liquid argon \cite{Lippincott:08}
and liquid neon \cite{Nikkel:08}. The LXe detector is made
of a cylinder of LXe viewed by two Hamamatsu R9869
 PMTs, as shown in Figure~\ref{fig:geant4_cell}. 
The PMTs  are specially designed for LXe applications. They 
have a bialkali photo-cathode and a quartz window with an aluminum
strip pattern on the photo-cathode. % for electrical conductivity to the photo-cathode. 
The two PMTs have a  quantum efficiency of 36\% for 
LXe scintillation light at 175~nm. The collection efficiency 
from the photo-cathode to the first dynode is about 70\%. The active 
LXe target is 5~cm in diameter and 2~cm in height and is surrounded 
by polytetrafluoroethylene (PTFE) for UV light reflection. The 
thickness of the PTFE is minimized (11.5~mm) to reduce neutron multiple scatters 
with surrounding materials. Two stainless steel mesh grids, each with 90\% 
optical transparency, are installed to apply  electric field to the
LXe. In dual phase mode, a third grid is added to apply a separate
electric field in the xenon gas region. 

The LXe detector is
located in an aluminum vacuum cryostat and cooled by a pulse-tube
refrigerator (PTR). The cryogenic system is described
elsewhere \cite{Ni:07}. The neutron generator is shielded by
30.5~cm~$\times$~30.5~cm~$\times$~30.5~cm water boxes  and two 10~cm $\times$ 30~cm $\times$ 30~cm polyethylene slabs to block and absorb neutrons that are emitted in directions other than toward the cryostat. 
The distance from the neutron generator to the center of the LXe was 76~cm. The distance between the center of the LXe and the center of the organic scintillator varied from 16 to 20~cm between runs. The scattering angle, defined by the position of the organic scintillator (Figure \ref{fig:setup}), was varied from 25 to 125 degrees to vary the associated recoil energy. 

During the single phase runs, the PMT waveforms  were recorded with an 8-bit oscilloscope, model TDS-5034B from Tektronix. The oscilloscope's logic gate was used to trigger the data acquisition system, 
triggering on triple coincidence (both $S1$ signals and the organic scintillator signal) in a 80~ns window.
  
\begin{figure}[htbp]
	\centering
		\includegraphics[width=3.25in]{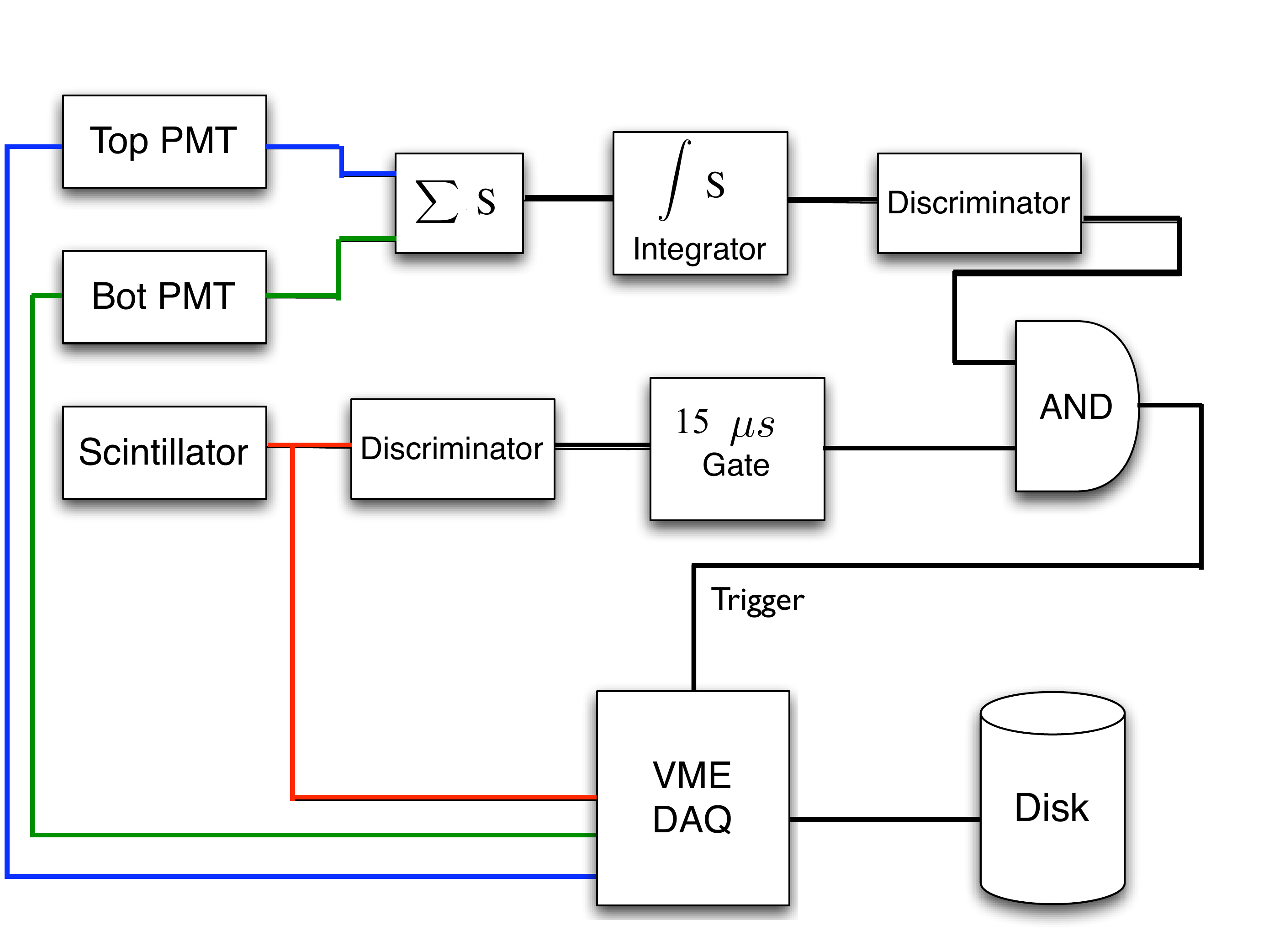}
	\caption{(Color online) 
	Trigger system used in the dual phase runs. The PMT signals are summed and integrated to select
	the $S2$ signals. The organic scintillator signal generates a 15 $\mu$s pulse. This pulse and the 
	$S2$ signals in coincidence trigger the acquisition system. For the single phase runs, triple 
	coincidence of the $S1$ signals and the organic scintillator signal triggers the acquisition system.}
	\label{fig:Dual_DAQ}
\end{figure}

For the dual phase runs, a VME 12-bit, 250 MS/s digitizer (CAEN V1720)  was used, because of its higher dynamic range. An external trigger for the VME digitizer was generated using external NIM modules.
In this mode the data acquisition system was triggered by the $S2$ signals. At a few scattering angles, data acquisition was instead triggered by the $S1$ signals to test for any systematic effects due to the trigger. 
For the $S2$ trigger,  shown in Figure \ref{fig:Dual_DAQ}, the PMT signals were added and integrated using a FAN IN/OUT and an integrator module. The summed and integrated signal then went to a discriminator  to select the $S2$ signals. The scintillator signal was sent through a discriminator to avoid small pulses. The output signal was used to generate a 15~$\mu$s pulse with the help of a gate generator. This pulse in coincidence with the $S2$ pulse generated the trigger for the acquisition system. For the $S1$ trigger we reproduced the trigger system described above for single phase operation, but using NIM modules.
In both the single and dual phase runs, the data were recorded  at a rate of $\sim$5~Hz.

\begin{figure}[htbp]
	\begin{center}
		\includegraphics[width=5.5cm, angle=90]{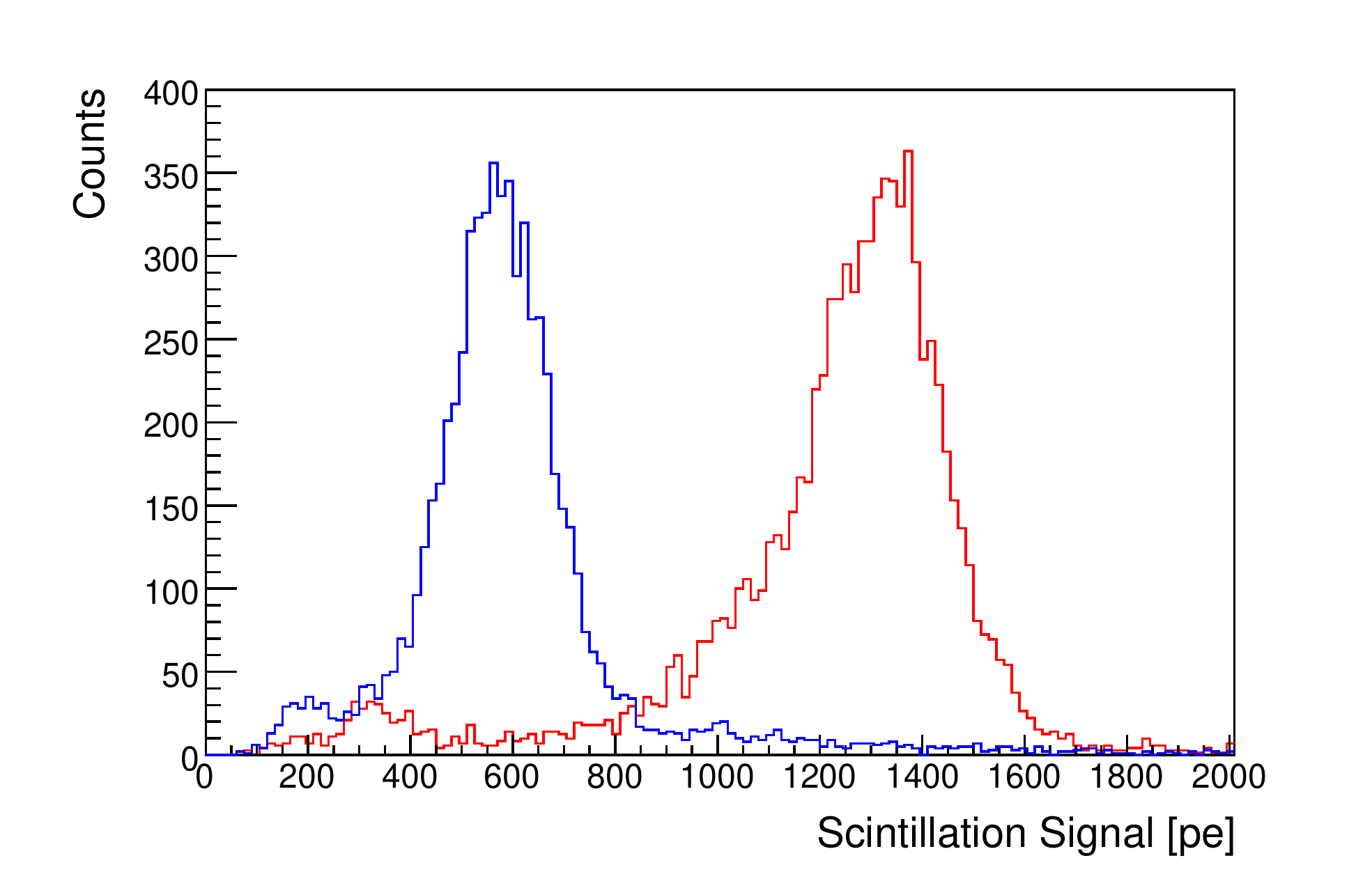}
		\caption{(Color online) Scintillation light yield for 122~keV gamma rays in the  
		LXe detector running in single phase.
		The left histogram was taken at 1.5~kV/cm yielding $4.8 
		\pm 0.1$~pe/keVee and a resolution ($\sigma/E$) of 18\%. The right 
		histogram was taken at 0.0~kV/cm yielding $10.8 \pm 0.1$~pe/keVee 
		and a resolution of 8.8\%.}
		\label{fig:co57}
	\end{center}
\end{figure}

Throughout the runs, periodic calibrations were performed to test the stability of the PMTs and measure the purity of the LXe.
The gains of the two PMTs were measured from the single photoelectron 
(pe) spectra by using light emitted from a blue LED located inside 
the LXe detector. The energy scale is calibrated using 
122~keV gamma rays from a $^{57}$Co source located outside the 
cryostat. The scintillation signal yields for the 122~keV gamma rays 
in LXe in the single phase detector are  $10.8 
\pm 0.1$ pe/keVee (keV in electron equivalent) at zero field and $4.8 
\pm 0.1$ pe/keVee at 1.5 kV/cm (Figure~\ref{fig:co57}). 
After adding a third grid and PTFE spacer and then removing some LXe to run the dual 
phase mode, the light yield drops to $9.5 \pm 0.2$~pe/keVee at zero 
field and $4.3 \pm 0.1$~pe/keVee at 1.0~kV/cm drift field. The 
light yields from the 122~keV events were monitored over the entire 
period of the measurements finding an additional fluctuation less than 3\%. 

\begin{figure}[htbp]
	\centering
		\includegraphics[width=5.5cm, angle=90]{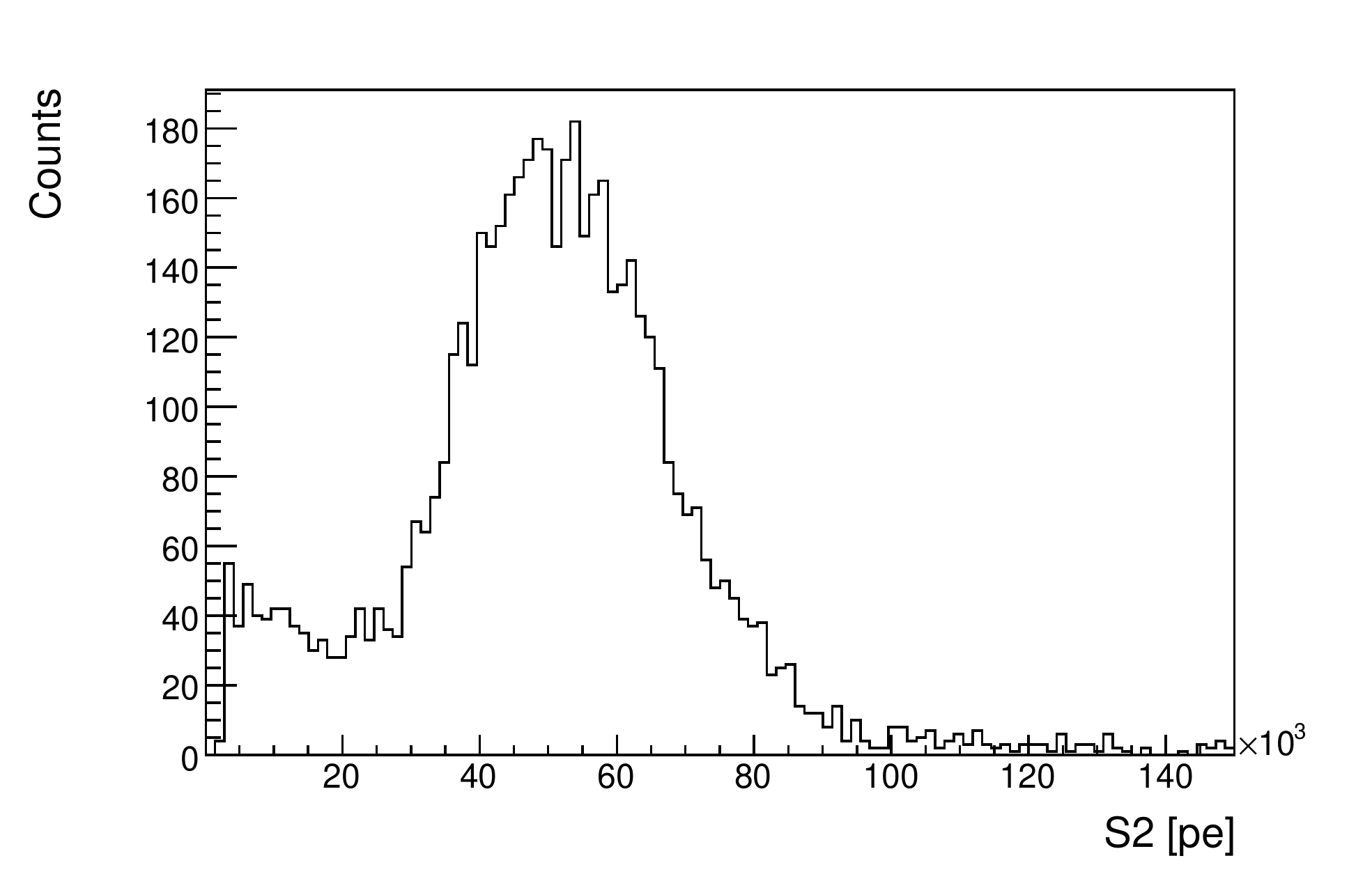}
	\caption{$S2$ spectrum for 122~keV gamma rays in the dual phase detector.}
	\label{fig:co57_s2}
\end{figure}

During the single phase runs, the LXe purity was monitored by 
measuring the stability of the scintillation signal yield from 122 keV
gamma rays. During the dual phase runs, the purity was also monitored
by measuring the electron lifetime, $\tau$, found by fitting the $S2$ 
spectrum with $S2 = S2_0 \exp{[-dt/\tau]}$ where $dt$ is the 
electron drift time, measured as the time between the $S1$ and $S2$
signals. For the data presented here, electron lifetime was 
greater than 40~$\mu$s and continuously improved over the course of the experiment. 
By the end of the experiment, the electron lifetime
was 90 $\mu$s.  
The total drift time for events at the cathode grid is  12 $\mu$s.
The $S2$ signals were corrected for the electron
lifetime. Figure \ref{fig:co57_s2} shows a typical $S2$ spectrum 
for 122~keV gamma rays taken during the dual phase runs.

\section{Finding  \lefftitle }
\label{sec:analysis}

To obtain the \leff~ value for each experimental setup we take the following steps. First, apply a set of cuts to reduce uninteresting events such as noise events, gamma ray scatters, neutron inelastic scatters and multiple elastic scatters. Next, the energy spectrum is obtained based on the light yield measurements and fitted to a  spectrum predicted through Monte Carlo simulation. The following subsections explain these steps.

\subsection{Data Analysis}
\label{sec:data_analysis}
\begin{figure}[htbp]
	\begin{center}
	$\begin{array}{c}
			\includegraphics[width=5.5cm, angle=90]{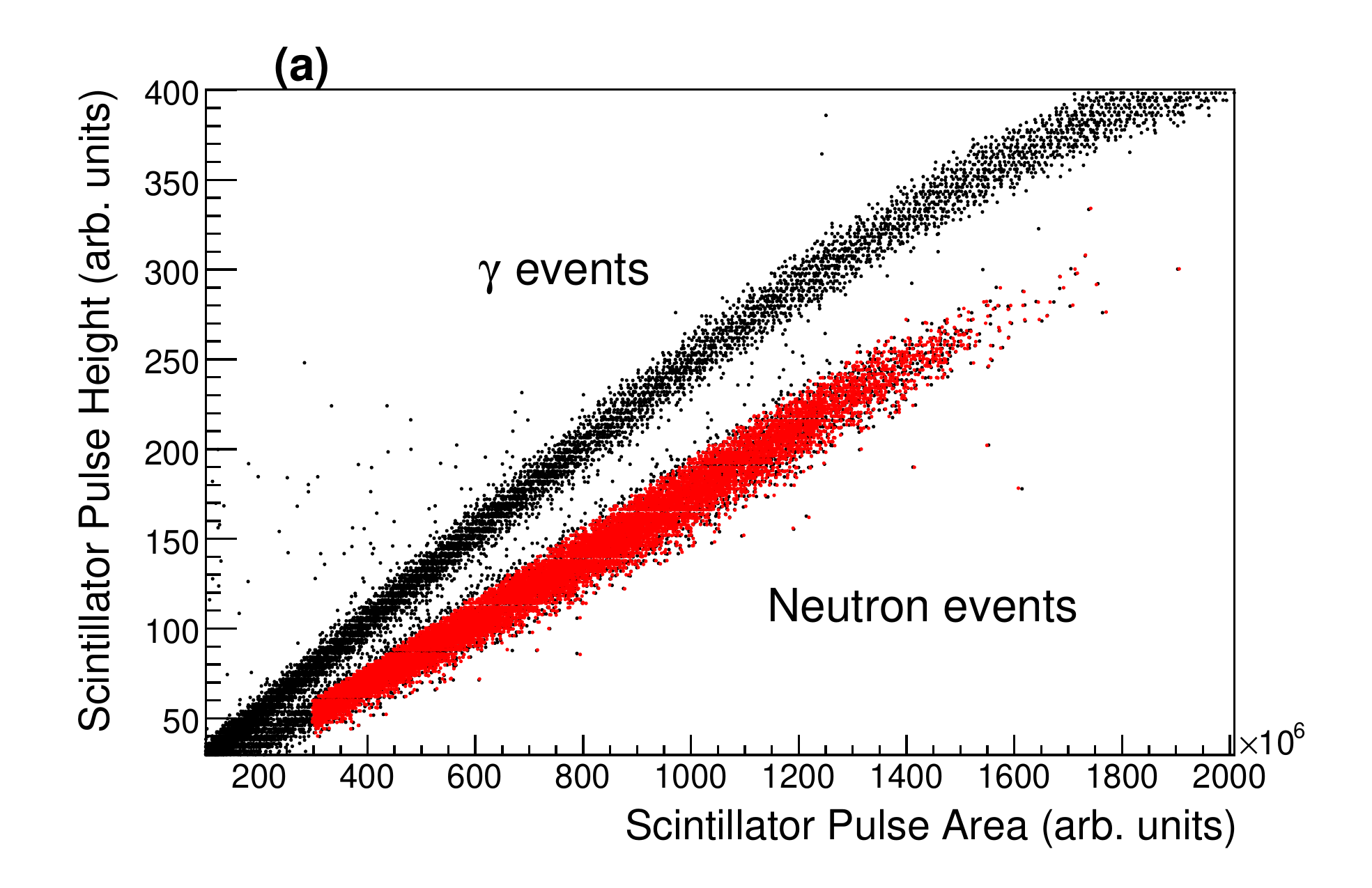} \\
			\includegraphics[width=5.5cm, angle=90]{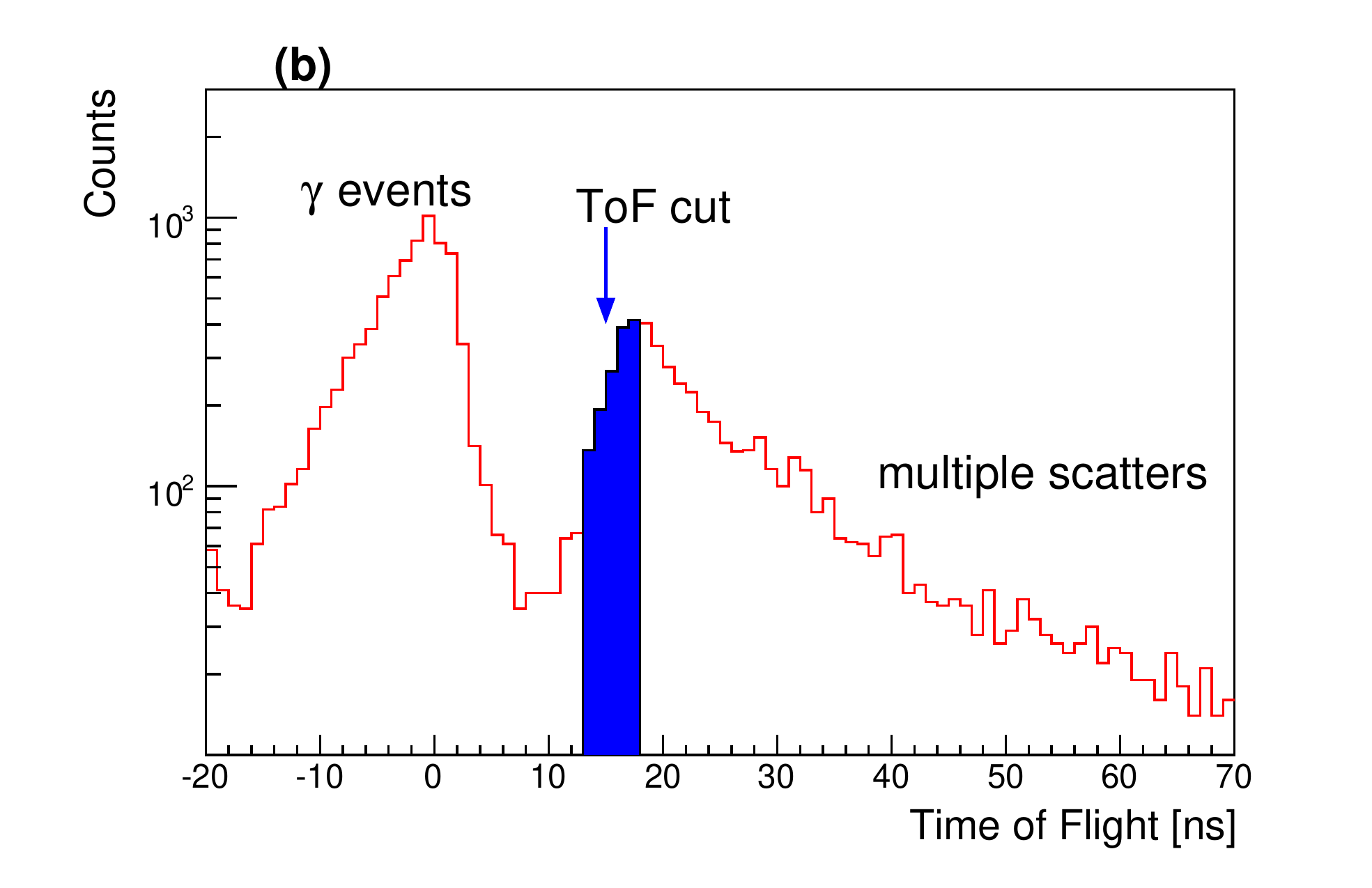} \\
	\end{array}	$
		\caption{(Color online) Two cuts used in the analysis to select single scatter 
		nuclear recoil events. \textbf{(a)} pulse shape discrimination in 
		the organic scintillator. \textbf{(b)} neutron time of flight 
		between the LXe and the organic scintillator. 
		Both plots show experimental data for 56~\kevr~run.}
		\label{fig:psdtof}
	\end{center}
\end{figure}

To remove uninteresting events two sets of cuts were used. The first set 
removes noise events and events outside the energy window of interest. 
The second set consists of two  cuts  used to select single elastic
nuclear recoil events (Figure~\ref{fig:psdtof}):
\begin{itemize}
	\item The first cut is based on 
	the pulse shape discrimination (PSD) of the organic scintillator. The 
	PSD is 
	 based on the relation between pulse height and pulse area. 
	This  is an effective way to separate the neutron events from gamma 
	events in the organic scintillator.
	\item The second cut uses the fact that 
	neutrons take a longer time to travel from the LXe detector to the 
	organic scintillator than gamma rays. The cut uses the time of flight 
	(ToF) to remove events triggered by gamma rays or
	accidental coincidence. The cut selects the first half of the ToF peak because the contribution from scatters other than single elastic ones is negligible, as determined from Monte Carlo simulations. 
\end{itemize}

\begin{figure*}[htbp]
	\centering		
	$\begin{array}{cc}
		\includegraphics[width=5.9cm, angle=90]{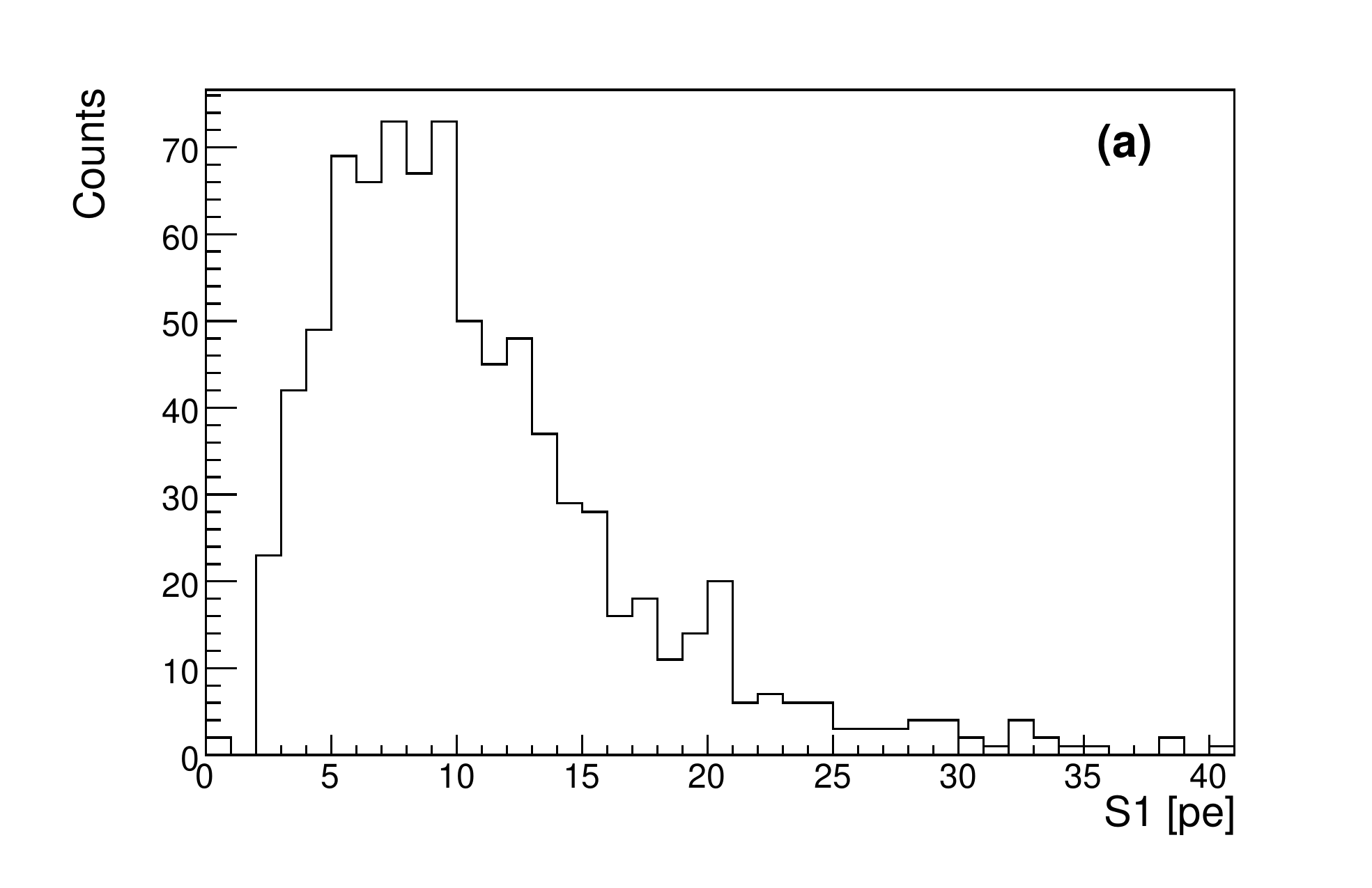} & 
		\includegraphics[width=5.9cm, angle=90]{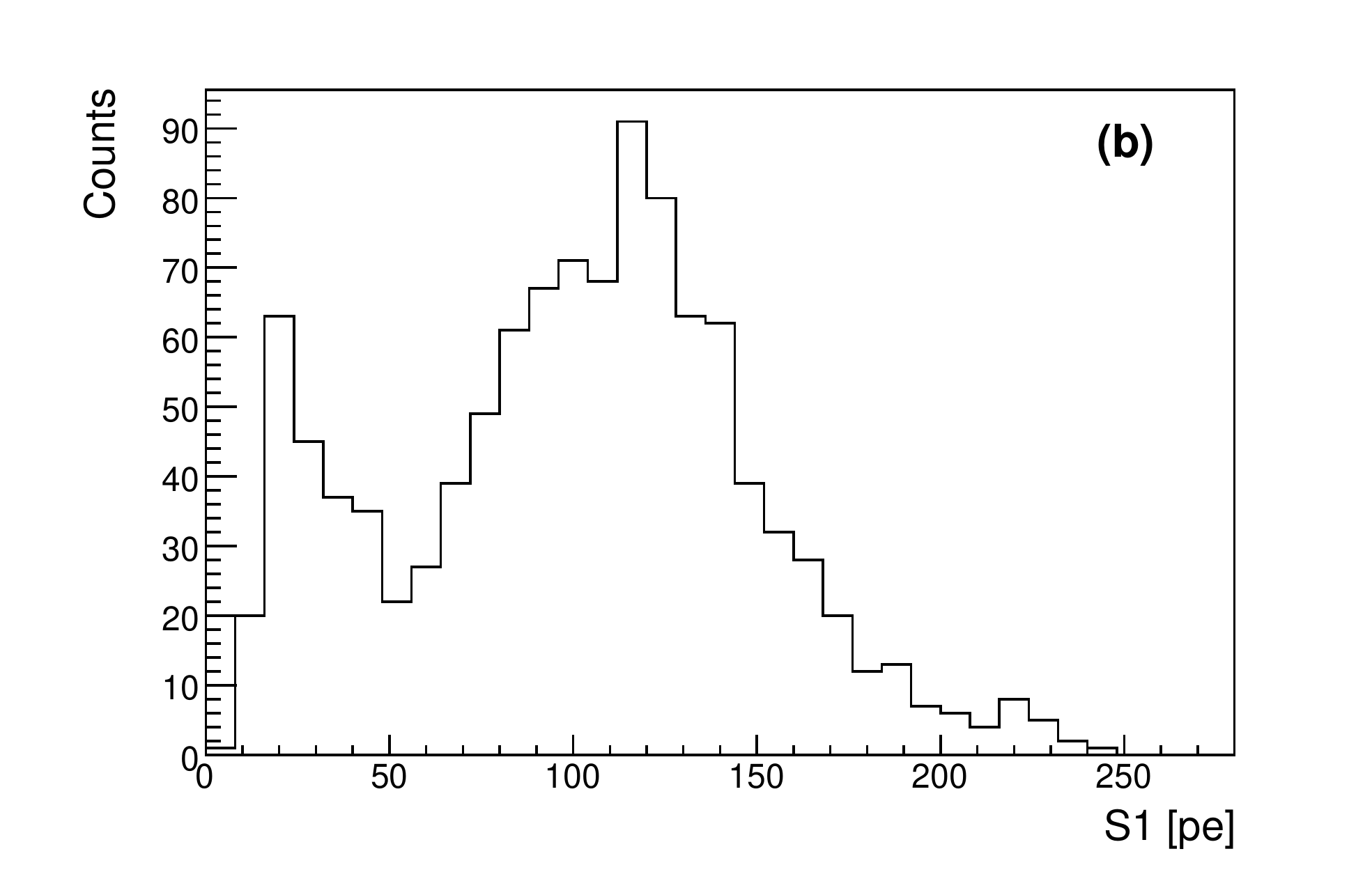} \\

		\includegraphics[width=5.9cm, angle=90]{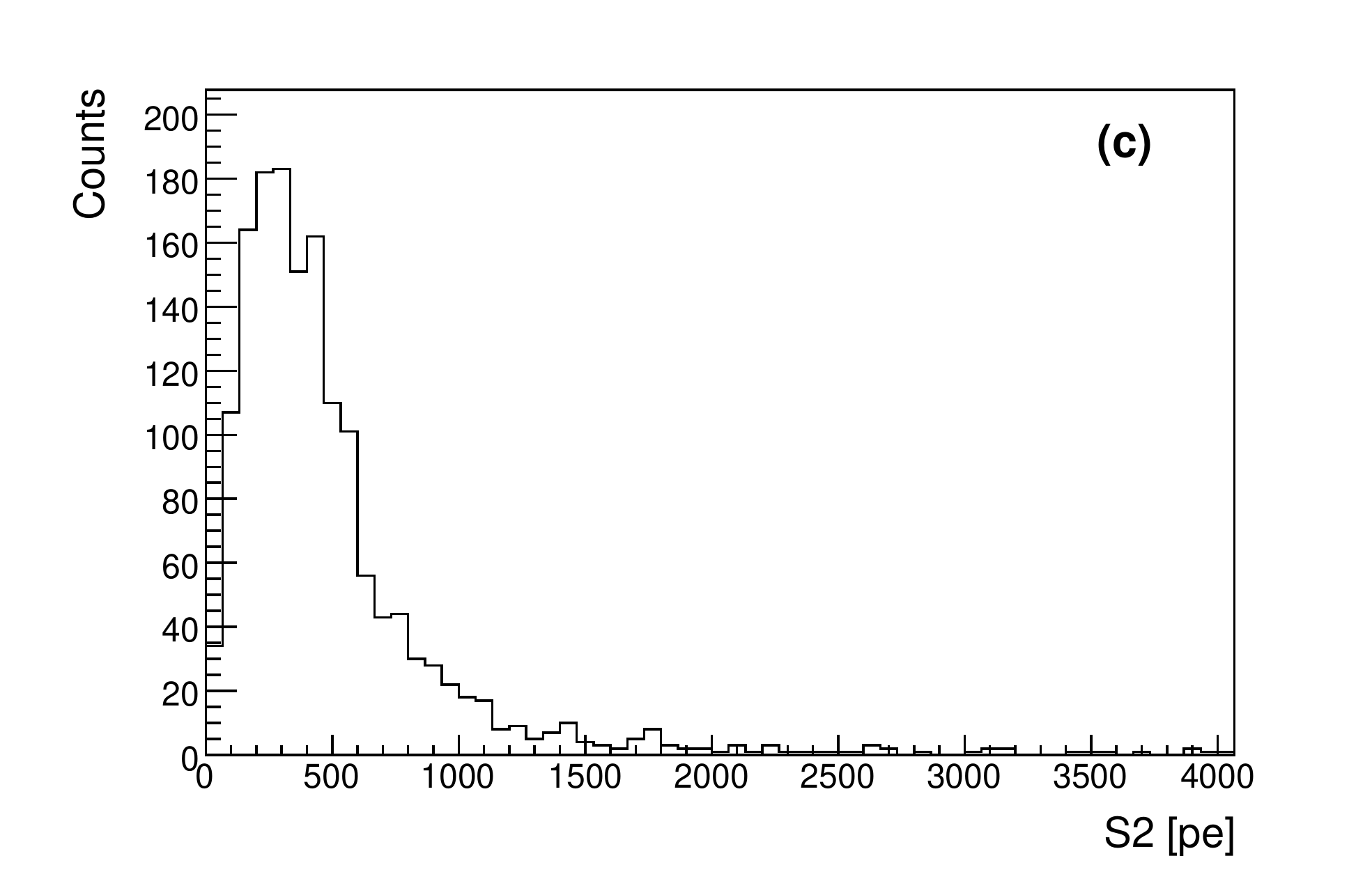} &
		\includegraphics[width=5.9cm, angle=90]{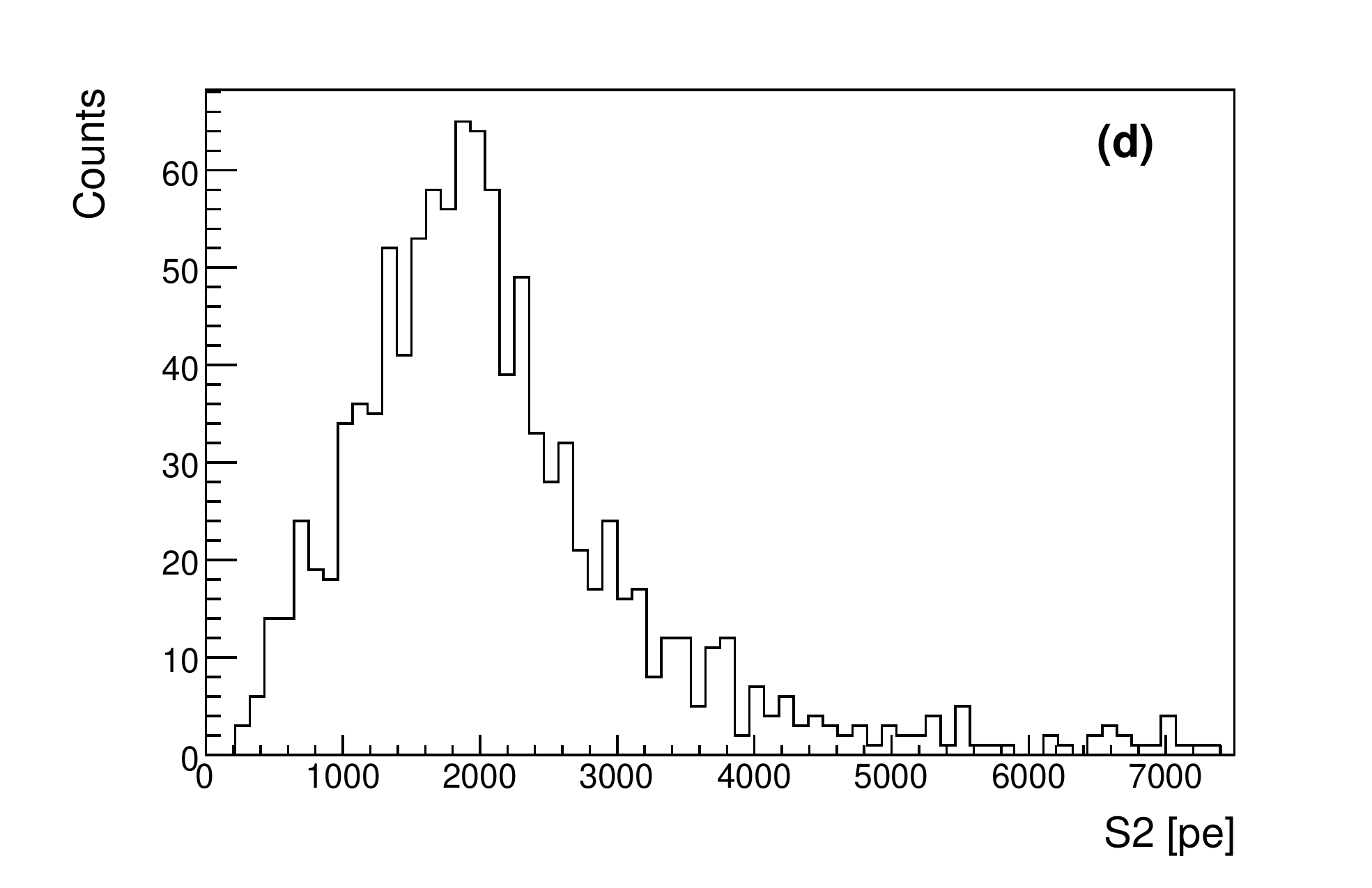} \\			
	\end{array}$
	
	\caption{$S1$ spectra in pe for \textbf{(a):} 6~\kevr~
	nuclear recoils at zero electric field and \textbf{(b):} 66~\kevr~ nuclear recoils at 1.5 kV/cm, 
	after applying the cuts.
	$S2$ spectra in pe for \textbf{(c):} 6~\kevr~
	nuclear recoils at zero electric field and \textbf{(d):} 66~\kevr~ nuclear recoils at 1.5 kV/cm.}
	\label{fig:S1S2_signals}
\end{figure*}

  The nuclear recoil energy distributions 
($S1$ signals) after these two cuts are shown in  
Figures~\ref{fig:S1S2_signals} \textbf{(a)} and \textbf{(b)} 
for 6 \kevr~ nuclear recoils at zero electric field and 66 \kevr~ nuclear recoils at 1.5 kV/cm, respectively. Similarly, Figures ~\ref{fig:S1S2_signals} \textbf{(c)} and \textbf{(d)} show $S2$ distributions for 6 \kevr~nuclear recoils at zero electric field and 66 
\kevr~ nuclear recoils at 1.5 kV/cm, after the cuts.
In the 56 and 66~\kevr~runs it is easy to separate the background tail from the single elastic signal, as shown in Figure \ref{fig:S1S2_signals} \textbf{(b)}. This tail has been removed ($S1 < 50$ pe) to generate the $S2$ spectrum in Figure \ref{fig:S1S2_signals} \textbf{(d)}, and to find the \leff~values.

\subsection{Monte Carlo simulation}
\label{sec:mc}
%\subsection{MC of neutron scattering} 
The detector's response to neutrons is modeled using a Geant4 
\cite{Geant4:03} simulation that takes into account the realistic setup 
of the experimental apparatus as described in Section~\ref{sec:exp}, 
including the water shield around the neutron generator, the 
polyethylene shield around the organic scintillator detector, the 
aluminum cryostat, the stainless steel cell, PTFE structure, PMTs, 
grids and LXe in the detector. The simulation stores the neutron 
scattering position, time, energy and type of events that deposit
energy in both the LXe detector and the liquid scintillator. 

For the simulations we follow the work by \cite{Sorensen:09} and used 
 the Xe(n,n)Xe scattering cross-sections from the updated ENDF/B-VII data, instead of the ENDF/B-VI cross-section used by default with the software. As discussed in \cite{Pignil:07}, a $\pm3\%$ uncertainty associated to the well depth parameter in the Optical Model Potential used to calculate the elastic cross-sections, leads to a conservative uncertainty which translates into an 
  uncertainty up to $\pm3\%$  in the nuclear recoil spectra. The latter is included when calculating the \leff~ systematic uncertainty.

\begin{figure}[htbp]
	\begin{center}
			\includegraphics[width=6.cm, angle=90]{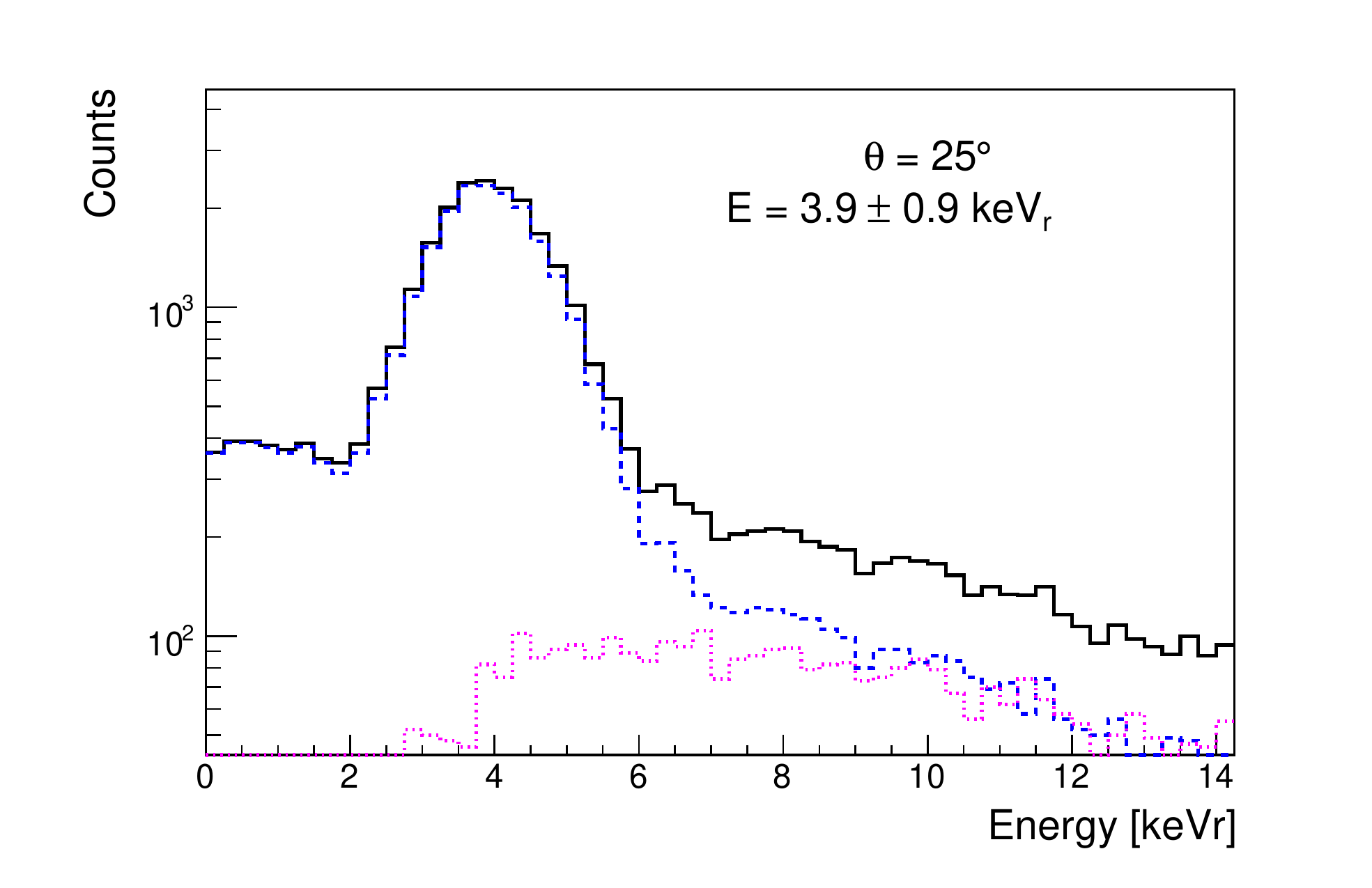}  
			\includegraphics[width=6.cm, angle=90]{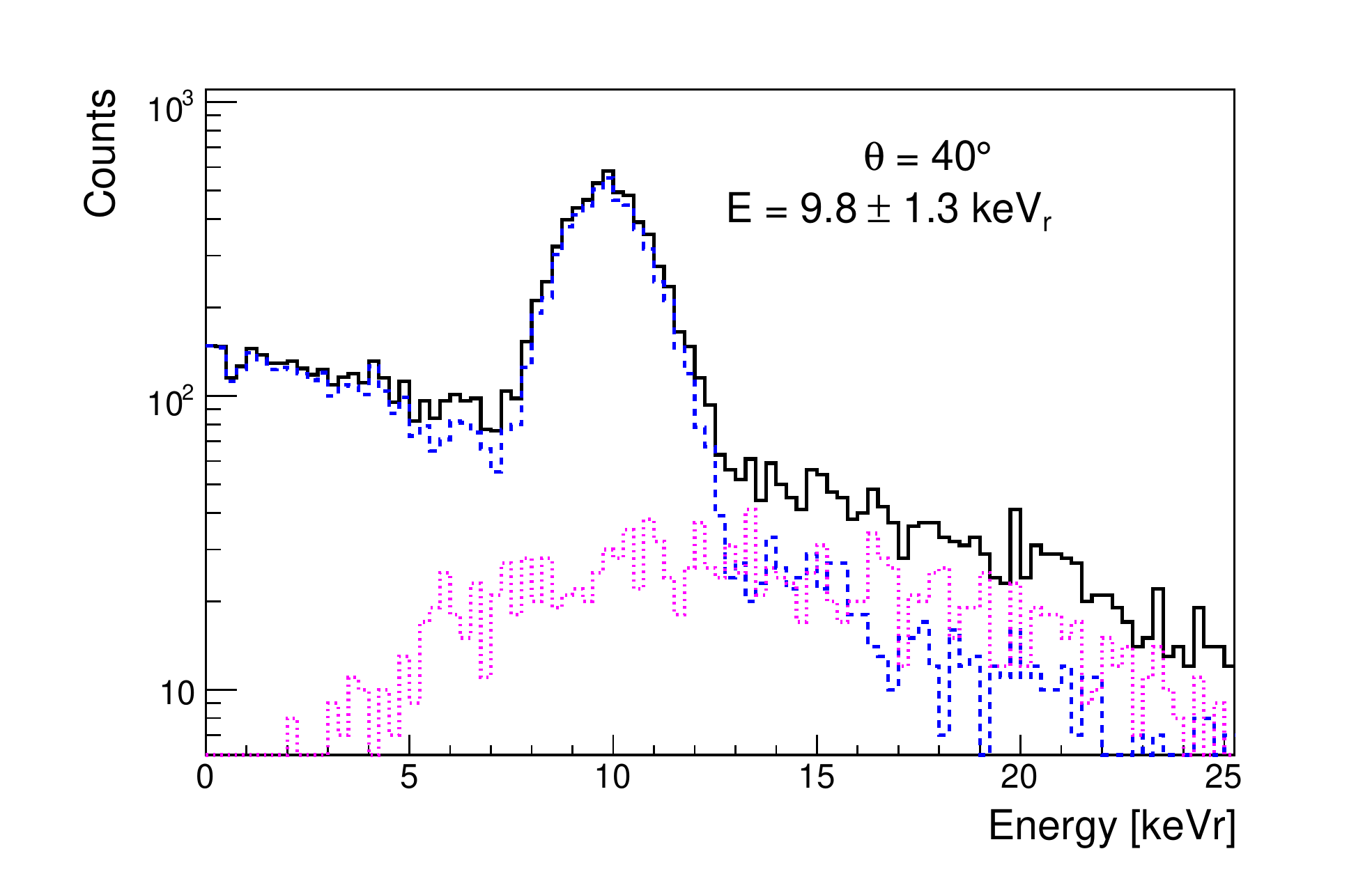} 	
			\includegraphics[width=6.cm, angle=90]{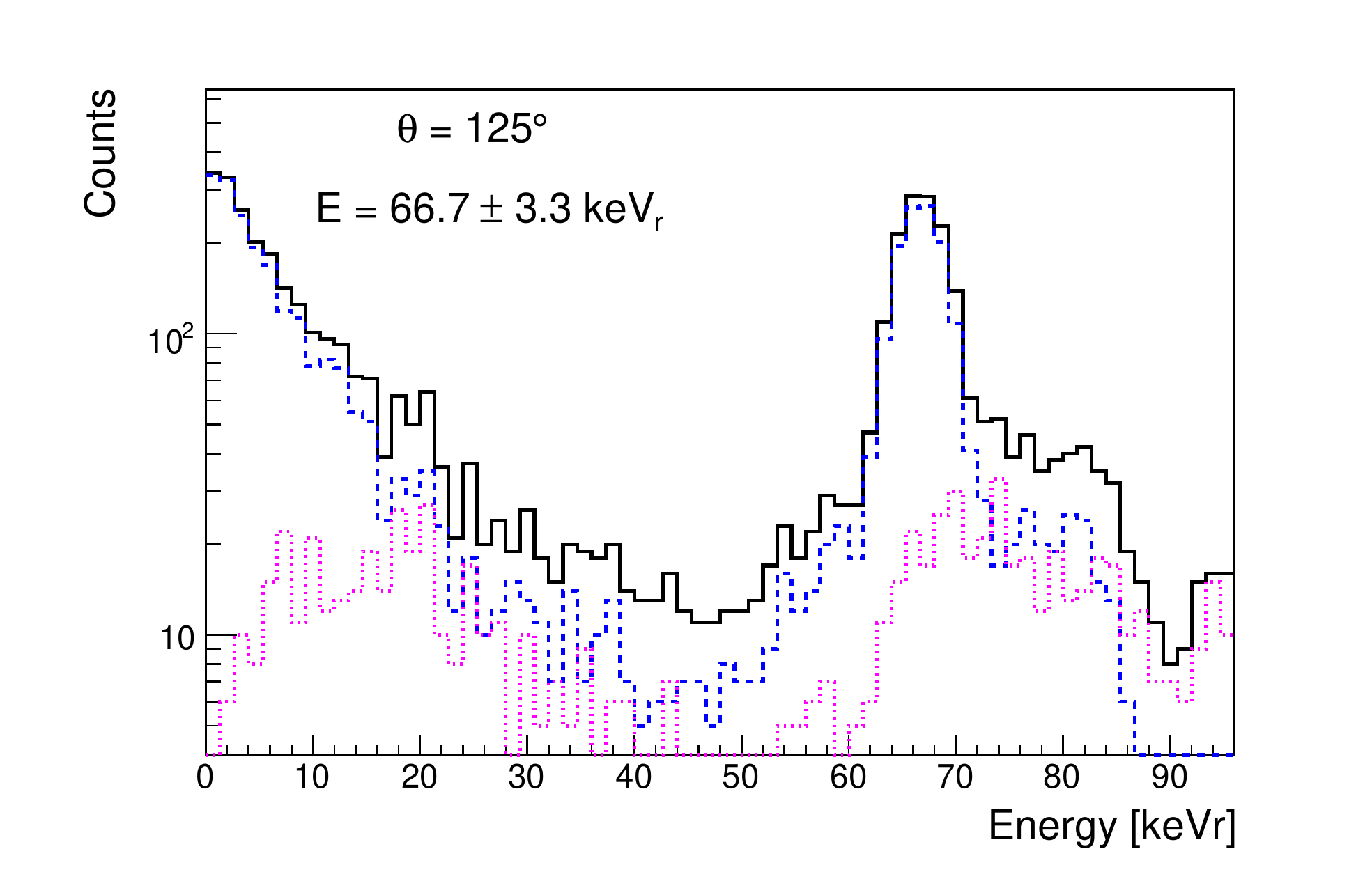} 
		\caption{(Color online) Simulated distribution of energy deposition in the LXe for 
			2.8~MeV neutrons that scatter and are tagged by the organic
			 scintillator, for three different angles, assuming the same ToF cuts
			 that were used in the experimental data. The solid line corresponds to all 
			types of events, the blue dashed line to single elastic nuclear recoils, 
			and the purple dotted line to multiple elastic scatters.  
			Multiple scatter events and inelastic events  are 
			negligible compared to the single scatter events. The energy and 
			uncertainty values indicated in the figure legends 
			are based on the mean and variance (one sigma) of
			 the peak for all events.}
		\label{fig:n_sim}
	\end{center}
\end{figure}

Neutrons can 
deposit energy in LXe via  elastic scattering, inelastic scattering, 
or a mixture of both. For most of the events that satisfy the ToF cut 
between the LXe and the liquid scintillator,  neutrons come directly 
from the neutron generator, make a pure single elastic scatter in LXe and 
reach the liquid scintillator. 
% Inelastic scatters with the LXe or the 
% surrounding materials are easily rejected  as they fall outside
 % the energy window of interest. 
Single elastic events 
give a peak (see Figure~\ref{fig:n_sim}) at the energy $E_r$ 
determined by the kinematics according to: 
\begin{equation}
\label{eq:kin}
E_r \approx E_n \frac{2m_n m_{Xe}}{(m_n + m_{Xe})^2} (1-\cos{\theta})
\end{equation}
where $E_n$ is the incoming neutron energy (2.8~MeV), while $m_n$ and 
$m_{Xe}$ are the masses of the neutron and the Xe nucleus, and $\theta$ is
the scattering angle. The spread of the energy deposition peak is 
from the  spread of $\theta$ due to the geometric width of the LXe
detector,  the width of the organic scintillator and the distance between 
them.

Some of the neutrons from the generator may scatter first in other 
materials (e.g. PTFE) before entering the LXe detector. Neutrons may also 
scatter more than once in the LXe detector. These ``multiple-scattering"
events have a variety of scattering angles and, therefore, deposit a wider range
of energies than the single elastic scatters. This contributes to the background tail under the 
pure single scattering peak. 
Through minimization of non-active material, the geometry of the detector was designed 
to reduce background from multiple scattering events, especially for runs performed 
at low scattering angles.
Inelastic scattering events are very few and 
make negligible contribution to the energy spectrum (less than 1\%). 
A detailed description of the contributions of each background for 
each energy tested can be found in \cite{Manzur:09}.

\begin{figure}[htbp]
	\centering
		\includegraphics[width=6cm, angle=90]{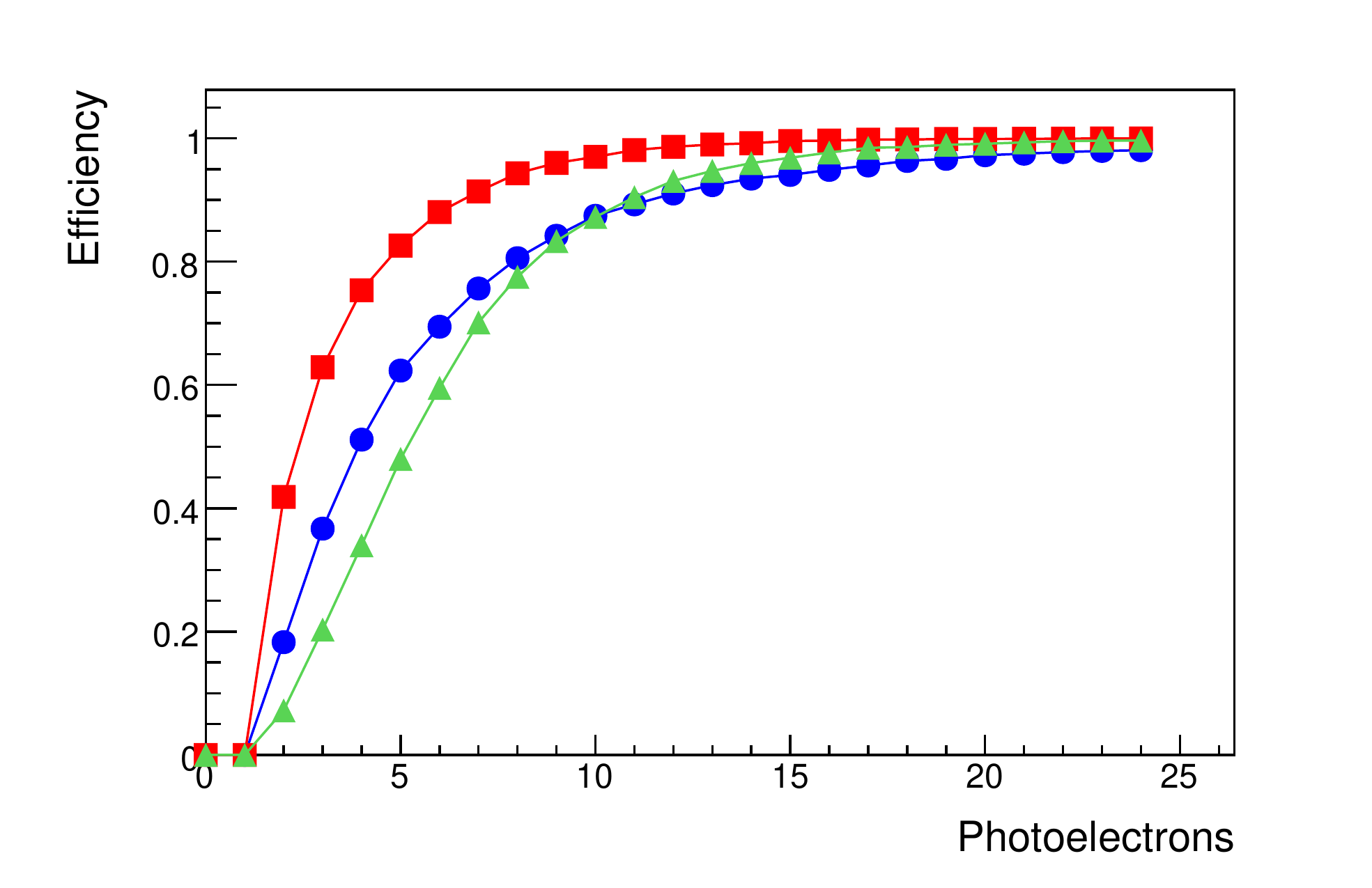}
	\caption{(Color online)  Detector triggering and software efficiencies for the 
	single (\textcolor{green}{\ding{115}}) and dual phase 
	(\textcolor{blue}{\ding{108}}) runs. The \textcolor{red}{\ding{110}} 
	points show the ideal case when the PMTs and the software have 100\% 
	efficiency for a single photoelectron.}
	\label{fig:s1_finding_eff}
\end{figure}

%\subsection{trigger efficiency}
Although the LXe detector gives a high scintillation yield, 
detecting and resolving the peak for low energy nuclear 
recoils is still challenging. 
The energy spread for nuclear recoils 
below 10 \kevr~ is dominated by statistical fluctuation of the 
photoelectrons in the PMTs. The trigger threshold,  the trigger 
efficiency and the $S1$-finding algorithm efficiency  must be taken 
into account in the analysis since not all events are effectively
detected. In our measurement, the trigger of the LXe signal is from
coincidence of the two PMTs with leading-edge discriminators. 
Effects of the trigger and software efficiencies are included in the 
Monte Carlo spectrum.
A realistic model of the trigger efficiency was determined by a Monte Carlo simulation that included
photon distribution into the two PMTs, quantum 
efficiencies,
statistical sampling of typical noise and single photoelectron  waveforms in the PMTs, 
the electronic trigger
thresholds, and the  $S1$-finding algorithm in the analysis software.
Figure~\ref{fig:s1_finding_eff} shows the overall trigger efficiency 
when combining these effects.

%\subsection{$\chi^2$ tests to find $L_{eff}$} 

\begin{figure*}[htbp]
	\centering		
	$\begin{array}{c@{\hspace{.0in}}c}
		\includegraphics[width=5.9cm, angle=90]{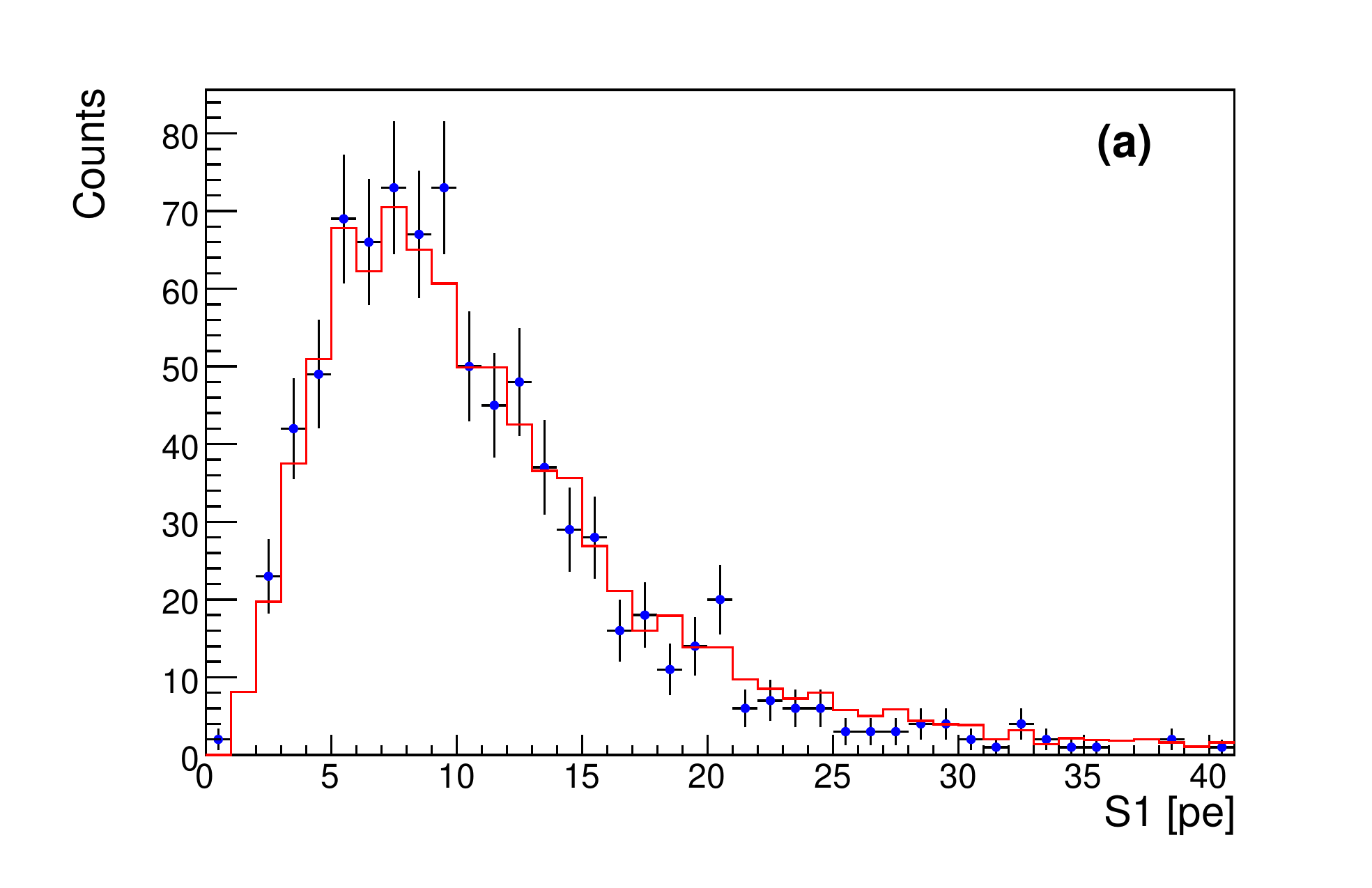} & 
		\includegraphics[width=5.9cm, angle=90]{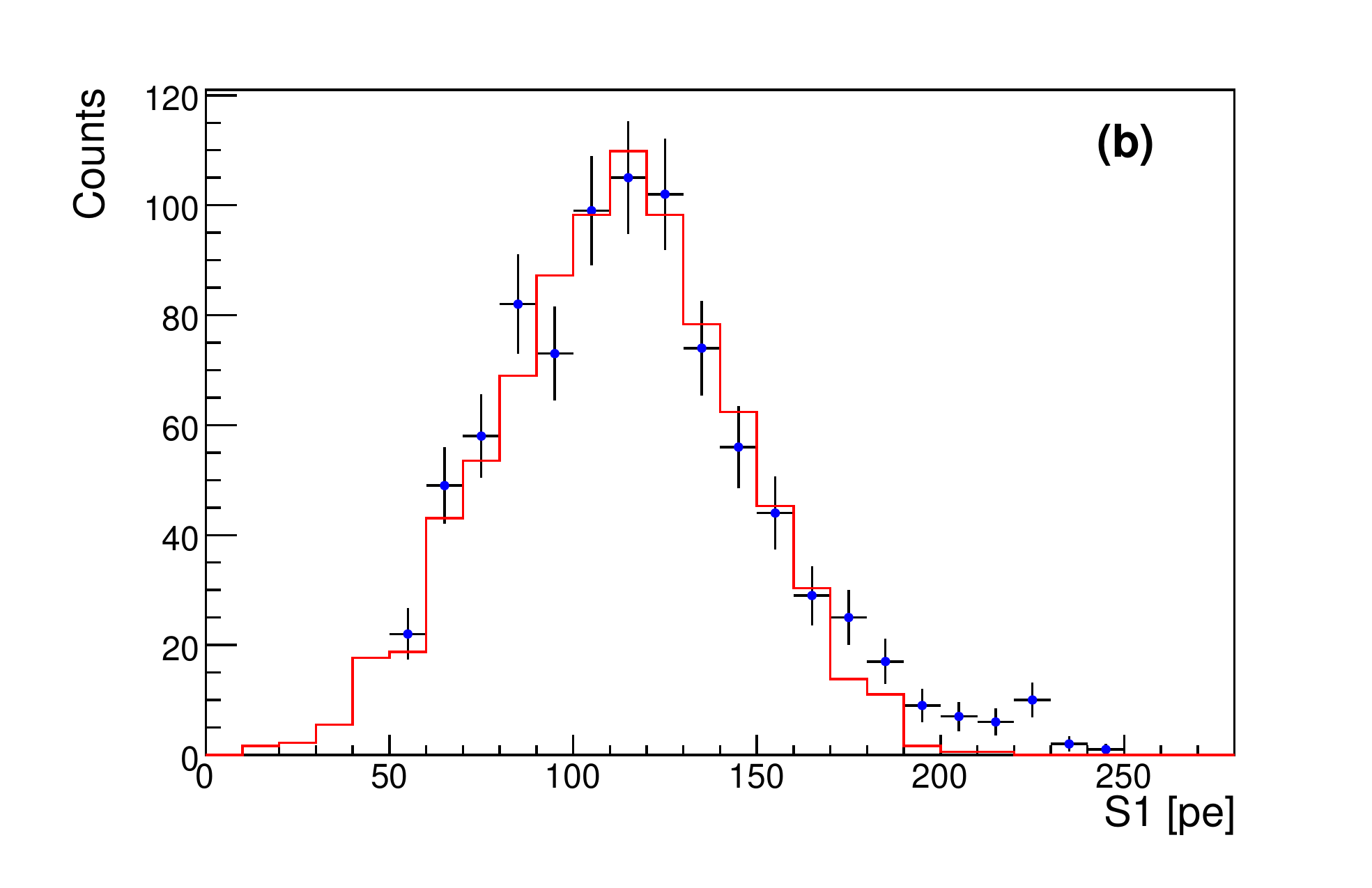} \\

		\includegraphics[width=5.9cm, angle=90]{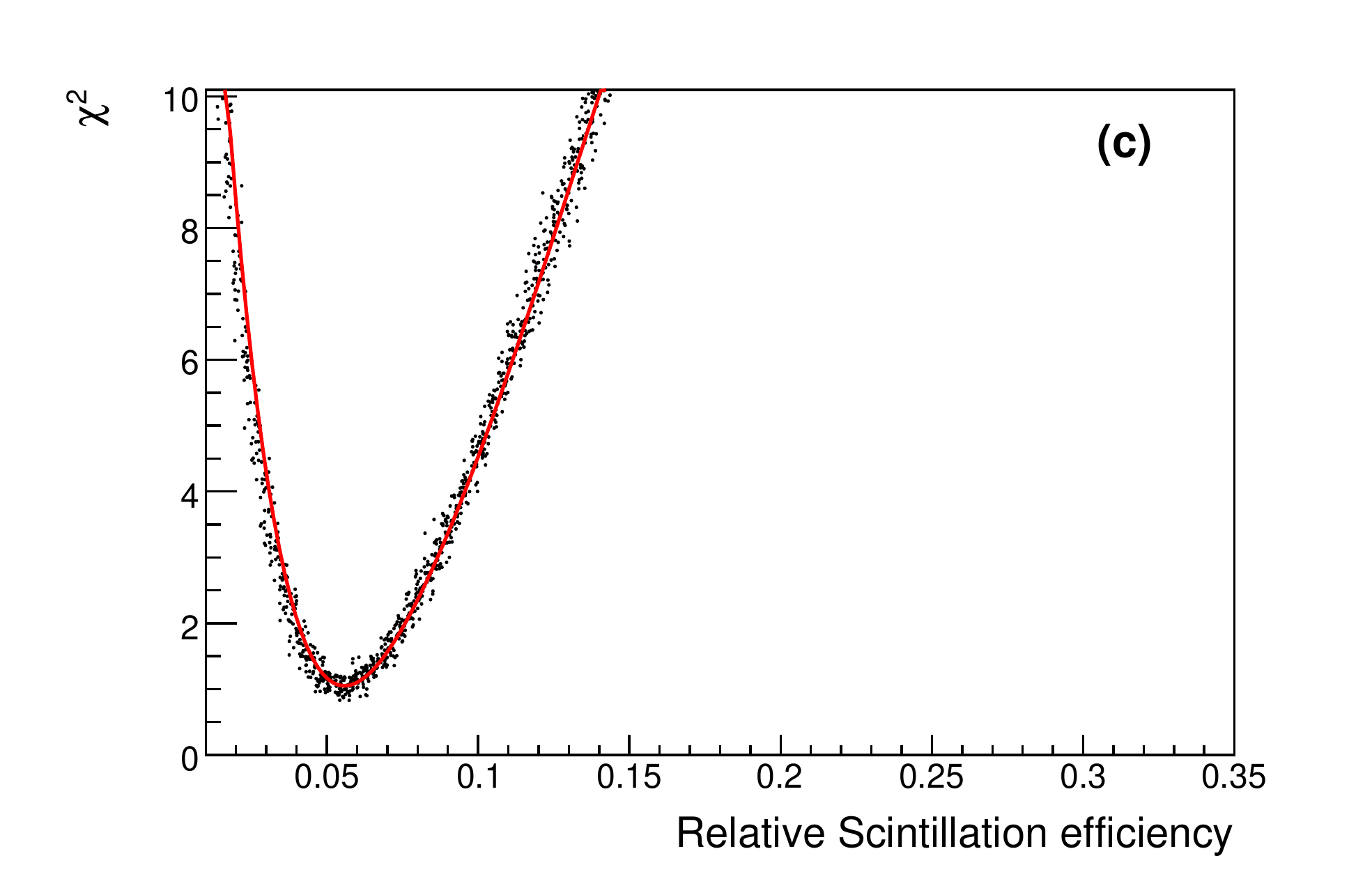} &
		\includegraphics[width=5.9cm, angle=90]{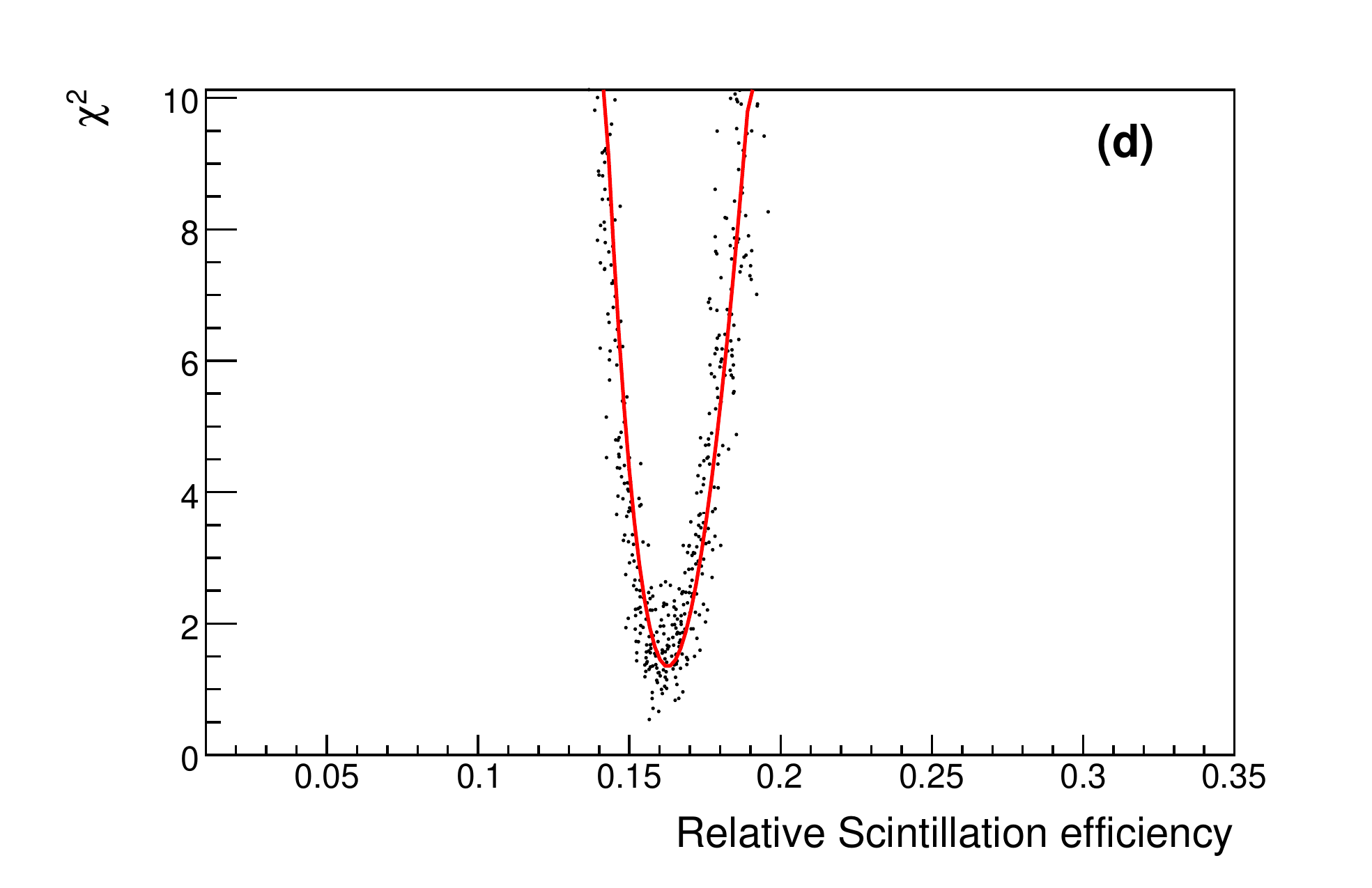} \\			
	\end{array}$	
	\caption{(Color online) $S1$ measured spectrum (data points) and Monte Carlo comparisons (solid line) for \textbf{(a)} 6~\kevr~ nuclear recoils at zero electric field and \textbf{(b)} 66 \kevr~ nuclear recoils at 1.5 kV/cm. The background tail ($S1 < 50$ pe) for the 66 \kevr~ run has been removed for the $\chi^2 $ test.
	Also shown are the $\chi^2$ vs \leff~ histograms for \textbf{(c)} 6~\kevr~nuclear recoil at zero electric field and \textbf{(d)} 66 \kevr~nuclear recoils at 1.5 kV/cm. The points show the different \leff~ values tested while the  curve is a fit to the points used to find the 1$\sigma$ errors on the \leff~ value. }
	\label{fig:mc_leff}
\end{figure*}

For each energy studied, the \leff~ value is found by comparing the 
Monte Carlo generated spectrum with the measured spectrum, using a  
$\chi^2$ test  according to equation \cite{PDG:06}:
\begin{equation}
\label{eq:chisq}
\chi^2 = \sum^N_{i=1} \frac{(n_i - \nu_i)^2}{\nu_i}
\end{equation}
where  $N$ is the 
total number of bins, while $n_i$ and $\nu_i$ are the measured and 
Monte Carlo generated number of events in each energy bin. 
To perform the $\chi^2$ test, the total number of events in the Monte Carlo 
spectrum was normalized to that of the measured spectrum.
The energy 
distribution (or $\nu_i$ at different energy bins) from the Monte 
Carlo spectrum varies with different input of \leff~ values. The 
best-fit \leff~ value is obtained by minimizing the $\chi^2$ parameter.
Figures~\ref{fig:mc_leff}	\textbf{(a)} and \textbf{(b)} show the data (points with error bars)
and the Monte Carlo spectrum (line) after minimization, for 6 \kevr~ nuclear recoils at zero electric field and 66 \kevr~ 
nuclear recoils at 1.5 kV/cm. These plots correspond to the $S1$ spectra shown in Figures~\ref{fig:S1S2_signals} \textbf{(a)} and \textbf{(b)}. 
Figures~\ref{fig:mc_leff}	\textbf{(c)} and \textbf{(d)} show the $\chi^2$ vs. \leff~ histograms. 
% The points show the different \leff~ values tested, the solid curve is a fit to the points used to
% find the 1$\sigma$ errors on the \leff value.

\section{Results}
\label{sec:results}
\begin{figure}[htbp]
	\begin{center}
		\includegraphics[width=5.5cm, angle=90]{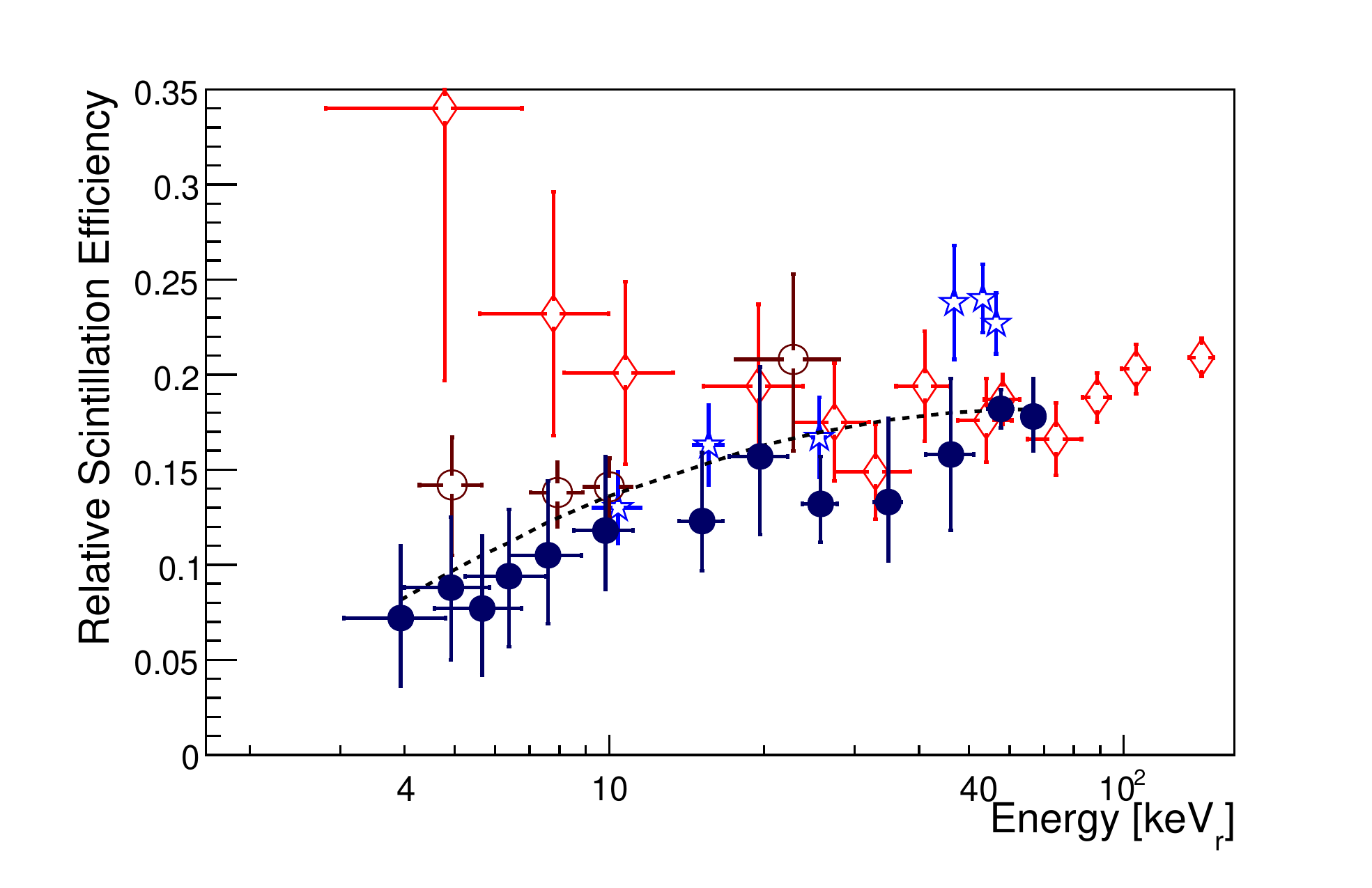}
		\caption{(Color online) Scintillation efficiency for nuclear recoils relative to 
		that of 122 keV gamma rays in LXe at zero field, comparing 
		this work (\textcolor{DarkBlue}{$\medbullet$}) to previous 
		measurements from Arneodo (\textcolor{black}{$\bigtriangleup$}) 	
		\cite{Arneodo:00}, Akimov (\textcolor{DarkGreen}{$\Box$}) 
		\cite{Akimov:02}, Aprile (\textcolor{blue}{\ding{73}})
		\cite{Aprile:05}, Chepel 
		(\textcolor{red}{$\Diamond$}) \cite{Chepel:06} and Aprile (\textcolor{DarkRed}{$\bigcirc$})\cite{Aprile:08}. Also shown is the 
		theoretical model (dashed line) explained in Section~\ref{sec:model}, 
		which includes the Lindhard factor, an electronic quenching 
		due to bi-excitonic collisions and the effect  of escaping electrons.}
		\label{fig:leff_result}
	\end{center}
\end{figure}

\begin{figure}[htbp]
	\centering
		\includegraphics[width=5.5cm, angle=90]{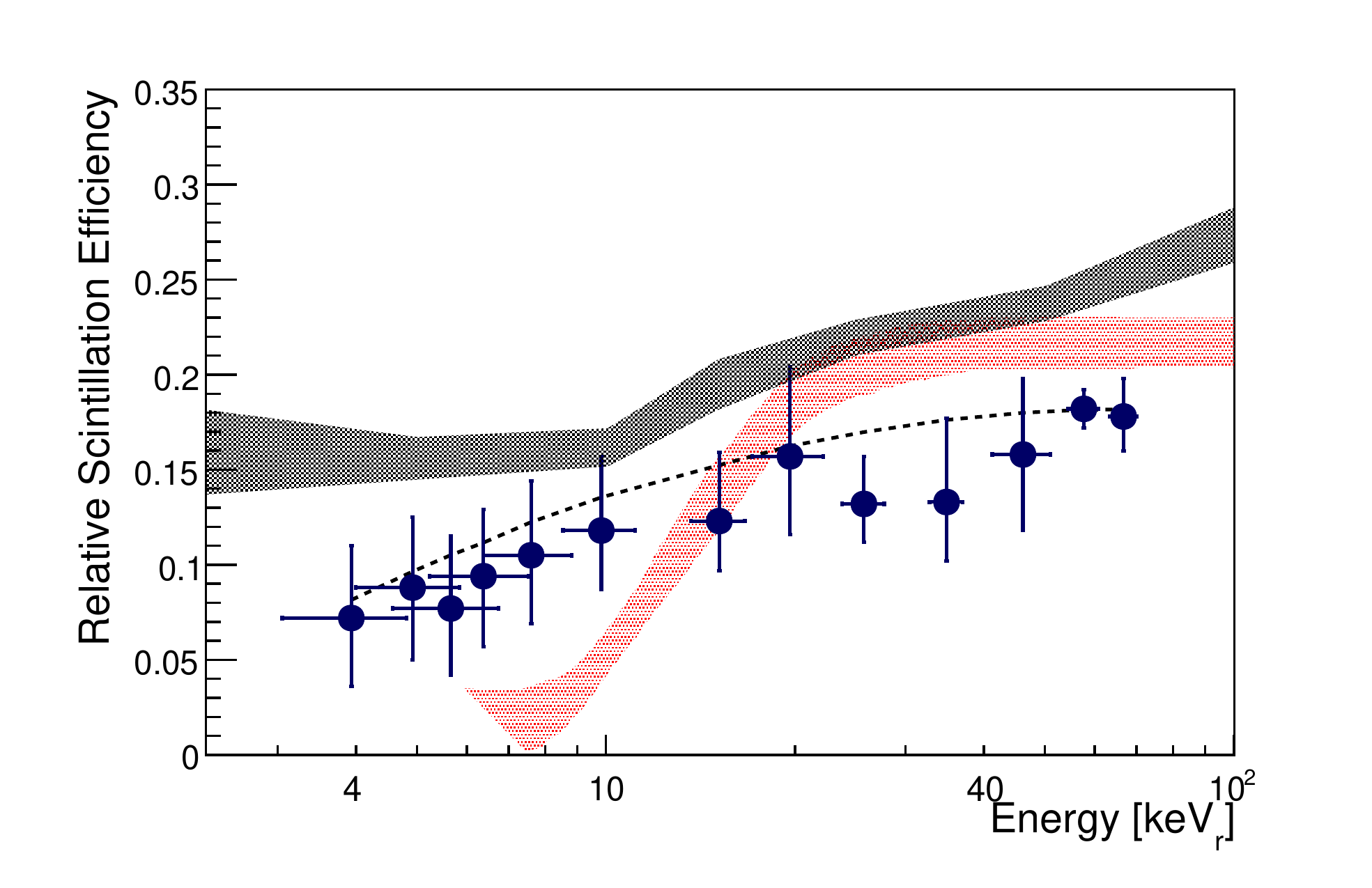}
	\caption{(Color online) Scintillation efficiency for nuclear recoils measured in 
	this work (\textcolor{DarkBlue}{$\medbullet$}) and the theoretical 
	model (dashed line) compared to the scintillation efficiency found 
	from the neutron calibration data by the XENON10 (top shaded area) \cite{Sorensen:09}
	and the ZEPLIN-III (bottom shaded area) \cite{Lebedenko:08} collaborations.}
	\label{fig:results_finalMC}
\end{figure}

%\subsection{uncertainty}
The largest  uncertainties in \leff~ derive from the $\chi^2$ analysis
and from the energy resolution applied to the Monte Carlo spectrum. 
The energy resolution is a combination of Poisson fluctuations in the PMTs, 
 the PMT gain fluctuations, the 
geometry of the cell ($\sigma_{geo}$), the optical properties of the 
materials  and the intrinsic resolution of the  
LXe. The total energy resolution, $\sigma$, defined 
as the root mean squared of the terms mentioned, was measured to be 
$\sigma = (3.2 \pm 0.4) \sqrt{N}$ with $N$ the number of 
photoelectrons for 56 and 66 \kevr. This relationship was assumed for 
determining all \leff~ values. 
Because the geometry resolution is already included in the simulation, the Monte Carlo data
were convolved using an  energy resolution $\sigma' \equiv \sqrt{\sigma^2 
- \sigma_{geo}^2}$.
The gain fluctuations were measured from the calibration 
runs, while the error due to the optical properties was determined from an independent light 
simulation. After subtracting the geometry resolution, the overall uncertainty on $\sigma'$  is
estimated to be $\pm 1.0 \sqrt{N}$, and this is propagated through to compute
the systematic uncertainty in \leff. The total  errors are shown in Figure~\ref{fig:leff_result} and given in Table \ref{tab:leff_vals}.  Figures~\ref{fig:leff_result} and \ref{fig:results_finalMC} compare 
our \leff~ results with previous analyses. Since the data show no 
significant nuclear quenching due to the electric field, \leff~ is
computed using data from all of the different runs, regardless of
electric field.

%field quenching
An electric field applied to LXe will suppress electron-ion 
recombination and thus reduce the scintillation yield. This field 
induced quenching is significant for electronic recoils. For example, 
at 1~kV/cm, scintillation light yield from 122~keV gamma rays in 
LXe is reduced to 53\% of its value at zero field 
\cite{Aprile:05}. For nuclear recoils, a near-unity value of $S_n$
 has been measured at 56~\kevr \cite{Aprile:05, Aprile:06}.
Here we measured the field induced quenching for nuclear recoils as low in energy as 4 
\kevr. No significant field 
quenching was observed for any energy or electric field. The average 
field induced quenching factor, $S_n$, for 56 \kevr~nuclear recoils at an 
electric field of 0.73 kV/cm  is about 95\% as given in Table~\ref{tab:leff_vals}.

\begin{figure}[htbp]
	\centering
		\includegraphics[width=6.0cm, angle=90]{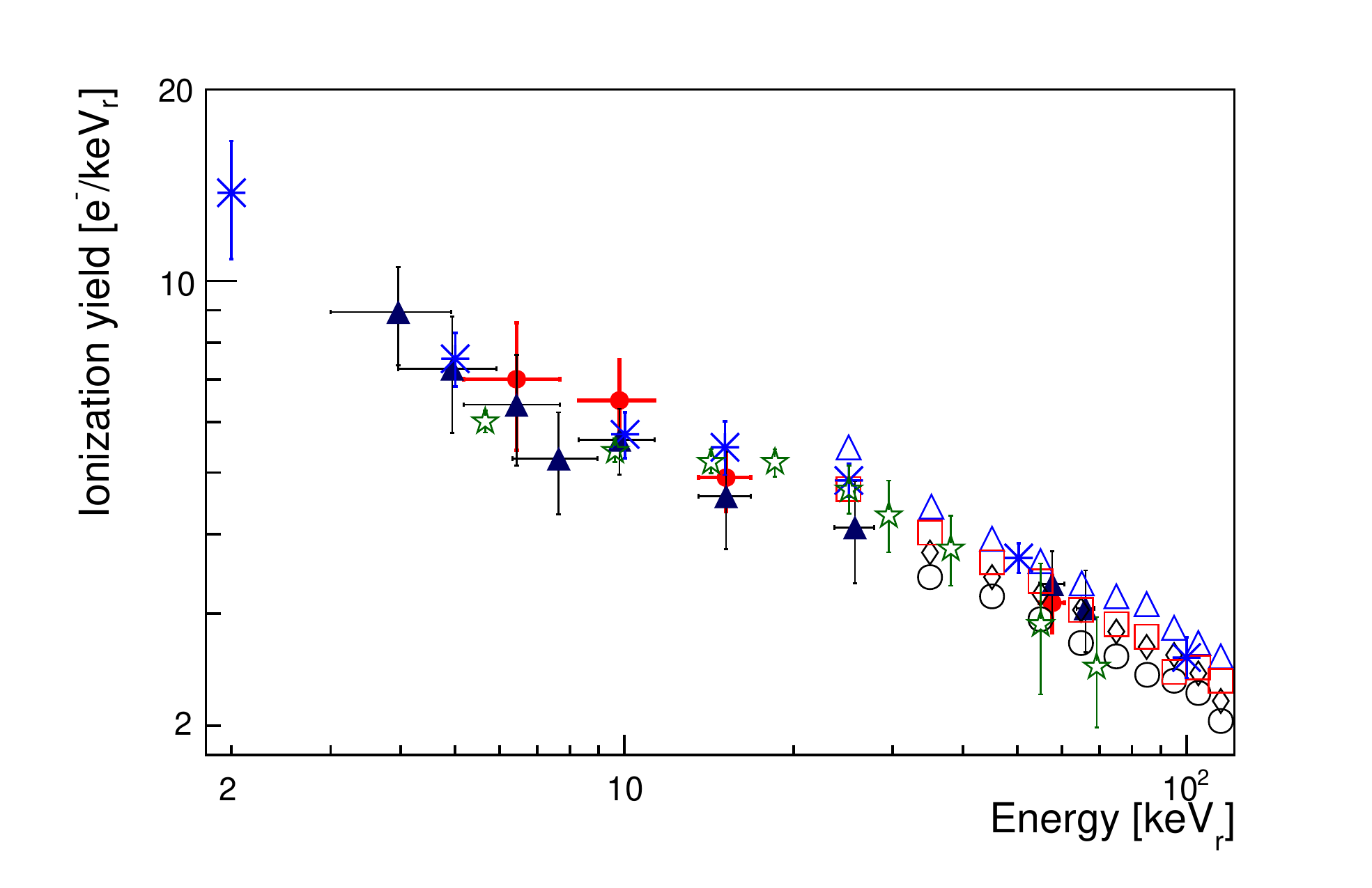}
	\caption{(Color online)  Ionization yield as a function of recoil energy. Shown are 
	the measured values in this work at 1.00~kV/cm 
	(\textcolor{DarkBlue}{\ding{115}}) and 4.00~kV/cm 
	(\textcolor{red}{$\medbullet$}), along with previously measured 
	values at 0.10~kV/cm (\textcolor{black}{$\bigcirc$}), 0.27~kV/cm (\textcolor{red}{$\Box$}), 2.00~kV/cm 
	(\textcolor{blue}{$\bigtriangleup$}) and 
	2.30~kV/cm(\textcolor{black}{$\Diamond$}) from \cite{Aprile:06}, 
	error bars  omitted for clarity. Also shown are the ionization yields calculated by comparing 
	the XENON10 nuclear recoil data and the Monte Carlo simulations \cite{Sorensen:09} 
	for single  elastic recoils at 0.73~kV/cm, using two different methods (\textcolor{blue}{\ding{83}} and  
	\textcolor{DarkGreen}{\ding{73}}).
	}
	\label{fig:QEQ0}
\end{figure}

From the dual phase data we can determine the ionization yield 
(number of electrons escaping recombination per unit recoil energy). 
This number is determined from the $S2$  peak position for the 
nuclear recoils (Figure~\ref{fig:S1S2_signals}\textbf{(d)} for example) and the 
number of photoelectrons per electron determined from the $^{57}$Co 
calibration runs. 
Figure \ref{fig:QEQ0} shows the energy dependence of the ionization 
yield measured in this work for 1.0~kV/cm and 4.0~kV/cm, as well as 
previous measurements \cite{Aprile:06} and the calculated values when comparing the XENON10 nuclear recoil data and Monte Carlo simulations \cite{Sorensen:09}.
The ionization yield errors shown in Figure \ref{fig:QEQ0} were derived from the
width of the $S2$ signals from nuclear recoils and from the $^{57}$Co calibrations. 
By comparing the dual phase runs
triggered by the $S1$ signals with the runs triggered by the $S2$ signals, we determined that there is no significant uncertainty in the $S2$ signals due to the trigger.

\begin{table*}
	\caption{\leff~and $S_n$ values  for the different 
	nuclear recoil energies studied. The third   column gives the \leff~values relative to 
	122 keV gamma rays. The first error is the statistical uncertainty, and the second error is the systematic uncertainty.} \label{tab:leff_vals}
\begin{center}
\begin{tabular}{cc|c|cccc}
 &  $E_r$ & \leff~  & $S_n$  & $S_n$ & $S_n$ & $S_n$ \\
$\theta$	&  [\kevr]	& at 0.0~kV/cm   & 0.73~kV/cm & 1.0~kV/cm & 1.5~kV/cm & 4.0~kV/cm\\
\hline \hline
125 &	66.7	$\pm$	3.3	&	0.178	$^{+	0.018 	+	0.010}_{-	0.016	-	0.009}$ 
	&	0.91	$\pm$	0.07	& 	1.11	$\pm$	0.09	&	0.88	$\pm$	0.06	&	-	\\ [1ex]
110 &	57.7	$\pm$	3.2	&	0.182	$^{+	0.009 	+	0.004}_{-	0.009	-	0.002}$ 
	& 	0.95	$\pm$	0.05	&	0.95	$\pm$	0.06	&	0.93	$\pm$	0.04	& 0.93	$\pm$	0.06	\\ [1ex]
95 &	46.1	$\pm$	4.9	&	0.158	$^{+	0.038	+	0.010}_{-	0.039	-	0.009}$ 
	&	0.97	$\pm$	0.08	& - &	0.82	$\pm$	0.08	& - \\ [1ex]
80 &	34.9	$\pm$	2.1	&	0.133	$^{+	0.042	+	0.014}_{-	0.029	-	0.012}$ 
	&	1.33	$\pm$	0.26	& - &	1.30	$\pm$	0.25	& - \\ [1ex]
67 &	25.7	$\pm$	2.0	&	0.132	$^{+	0.025	+	0.001}_{-	0.019	-	0.006}$ 
	&	0.95	$\pm$	0.06	&	0.91	$\pm$	0.12	& 0.95	$\pm$	0.07	& - \\ [1ex] 
58 &	19.6	$\pm$	2.6	&	0.157	$^{+	0.046	+	0.008}_{-	0.036	-	0.019}$ 
	&	0.70	$\pm$	0.06	& - &	1.03	$\pm$	0.13	& - \\ [1ex]
50 &	15.1	$\pm$	1.5	&	0.123	$^{+	0.030	+	0.019}_{-	0.023	-	0.014}$ 
	&	0.83	$\pm$	0.16	&	1.02	$\pm$	0.20	&	1.01	$\pm$	0.15	& 1.08	$\pm$	0.18\\ [1ex]
40 &	9.8	$\pm$	1.3	&	0.118	$^{+	0.027	+	0.029}_{-	0.022	-	0.022}$ 
	&	0.91	$\pm$	0.17	&	1.64	$\pm$	0.50 	&	0.98	$\pm$	0.18	& 1.62	$\pm$	0.45\\ [1ex]
35 &	7.6	$\pm$	1.2	&	0.105	$^{+	0.028	+	0.026}_{-	0.022	-	0.029}$ 
	&	0.79	$\pm$	0.28	&	1.06	$\pm$	0.30	&	0.79	$\pm$	0.28	& - \\ [1ex]
32 &	6.4	$\pm$	1.1	&	0.094	$^{+	0.027	+	0.023}_{-	0.022	-	0.029}$ 
	&	0.92	$\pm$	0.37	&	1.25	$\pm$	0.45	&	0.93	$\pm$	0.38	& 1.38	$\pm$	0.52\\ [1ex]
30 &	5.7	$\pm$	1.1	&	0.077	$^{+	0.028	+	0.027}_{-	0.022	-	0.026}$ 
	&	1.35	$\pm$	0.67	& - &	1.18	$\pm$	0.61	& - \\ [1ex]
28 &	4.9	$\pm$	0.9	&	0.088	$^{+	0.026	+	0.026}_{-	0.023	-	0.032}$ 
	&	1.16	$\pm$	0.45	&	1.34	$\pm$	0.50	&	0.87	$\pm$	0.35	& - \\ [1ex]
25 &	3.9	$\pm$	0.9	&	0.073	$^{+	0.034	+	0.018}_{-	0.025	-	0.026}$ 
	&	1.19	$\pm$	0.52	&	1.30	$\pm$	0.38	&	1.88	$\pm$	0.78	& - \\ [1ex]
\hline

\end{tabular}
\end{center}
\end{table*}

\section{Empirical model of \lefftitle}
\label{sec:model}
The data shown above reveal a relative scintillation efficiency that
decreases with decreasing energy. A suitable theoretical expression for
\leff~ in LXe can be written as the product of at least three components:

\begin{equation}
\label{eq:totqf}
\leffeq = q_\textrm{ncl} \cdot q_\textrm{esc} \cdot q_\textrm{el} 
\end{equation}

First is the Lindhard factor \cite{Lindhard:63}, 
$q_\textrm{ncl}$, which quantifies the larger fraction of energy dissipated
into atomic motion or heat in a nuclear recoil compared to that for
an electron recoil. As a function of recoil energy, $E_r$, the Lindhard factor 
can be written as
\begin{equation}
	\label{eq:qncl}
		q_\textrm{ncl} = \frac{k\cdot g(\varepsilon)}{1 + k \cdot g(\varepsilon)}
\end{equation}
where for a nucleus with atomic number $Z$ and mass number $A$, 
$k = 0.133 \cdot Z^{2/3} \cdot A^{-1/2}$, 
$g(\varepsilon) = 3.0 \varepsilon^{0.15} + 0.7 \varepsilon^{0.6} + \varepsilon$,
 where $\varepsilon$ is the reduced energy $\varepsilon = 11.5 \cdot E_r \cdot Z^{-7/3}$.

\begin{figure}[htbp]
	\centering
		\includegraphics[width=6cm, angle=90]{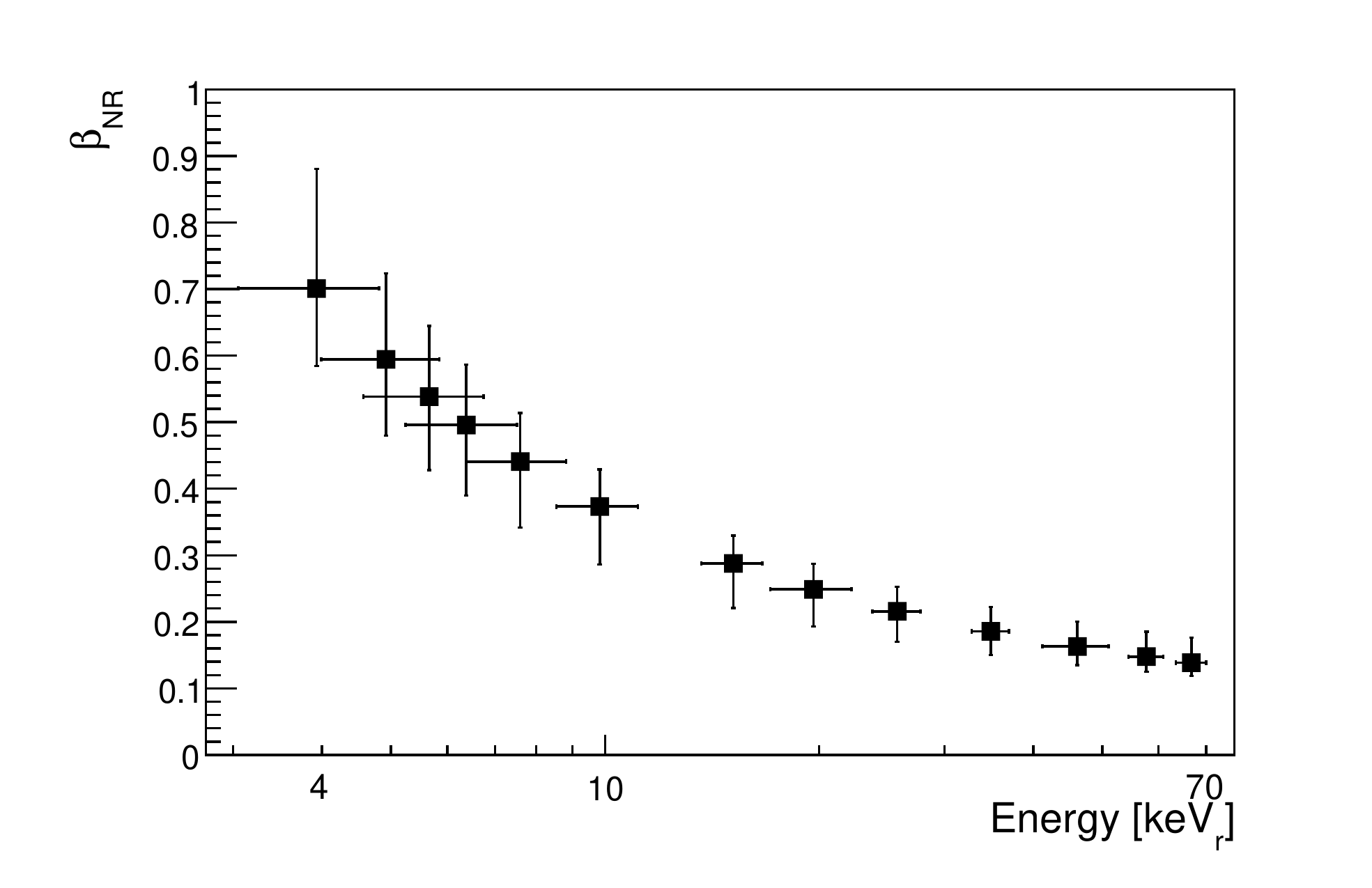}
	\caption{Fraction of escape electrons for nuclear recoils, $\beta_{\textrm{NR}}$,  as a function of 
	recoil energy used in this work.}
	\label{fig:Nesc_mine_kevr}
\end{figure}

The second term, $q_\textrm{esc}$, is the reduction of the scintillation light yield due to 
\emph{escaping electrons}. These  are electrons
produced by ionization that thermalize outside the Onsager radius and 
become free from recombination even in the absence of an electric field 
\cite{Doke:02}.  The effect of escaping electrons has been observed for 
electron recoils %, such as those from 122~keV gammas 
\cite{Doke:02} and
 has only recently been 
considered as a possible additional factor governing the total 
scintillation reduction for nuclear recoils in LXe 
\cite{Shutt:pri}. This is because of the surprisingly high ionization yield from 
nuclear recoils \cite{Aprile:06}. This factor can be expressed in 
terms of the ratio between the initial number of excitons 
and electron-ion pairs $\alpha \equiv N_\textrm{ex}/N_\textrm{i}$, and  the 
fraction of escape electrons over the total electron-ion pairs $\beta \equiv N_\textrm{esc}/N_\textrm{i}$.
\begin{equation}
q_\textrm{esc} = \frac{N_\textrm{ex} + N_\textrm{i} - 
N_\textrm{esc}}{N_\textrm{ex} + N_\textrm{i} - N_\textrm{esc}^{122}} 
=  \frac{\alpha +  1- \beta_{\textrm{NR}}}{\alpha + 1 - \beta^{122}}
\end{equation}
For this work we select 
$\alpha = 0.06$, measured for electron recoils in LXe \cite{Doke:02, Dahl:09-phd}.
$\beta^{122}$ is the fraction of escaping electrons for 122~keV
electron recoils in LXe and is calculated to be 0.31 based on the 
$^{57}$Co data from \cite{Aprile:06} and the method described in \cite{Doke:02}.
% LET (linear energy transfer) dependence of scintillation yield (see 
% Figure~4 in \cite{Doke:02}). 
$\beta_{\textrm{NR}}$ can be calculated based on 
the nuclear recoil ionization yield reported in \cite{Aprile:06}
and those measured in this work (Figure~\ref{fig:QEQ0}). The $\beta_{\textrm{NR}}$
values estimated for the different energies studied in this work are 
given in Figure \ref{fig:Nesc_mine_kevr}.

The last term in the \leff~model is the scintillation light quenched 
by  bi-excitonic collisions, $q_\textrm{el}$, as proposed by Hitachi 
\cite{Hitachi:05} to explain the much lower measured \leff~ values
 than are predicted by $q_{\textrm{ncl}}$ alone. 
Bi-excitonic collisions have the effect of two excitons producing a 
single photon instead of two photons. A recent paper \cite{Mei:08}  
extends the study of quenching due to bi-excitonic collision by 
including the varying quenching due to different stopping power for 
different energy recoils, as quantified by Birks' Law \cite{Birks:51}. Thus,
\begin{equation}
q_\textrm{el} = \frac{1}{1 + k_{B} \cdot \frac{dE}{dx}}
\end{equation}
We obtain $dE/dx$ values from SRIM \cite{SRIM:09} and fit the $k_B$ parameter to match  our \leff~ model to the data at 56~\kevr, finding $q_\textrm{el} = 0.65 $. Hitachi 
\cite{Hitachi:05} used $q_\textrm{el} = 0.68$  to match his model and the data 
at  60~\kevr.

\section{Discussion}
\label{sec:discussion}
The high energy points (56 and 66 \kevr) agree with previous \leff~ 
measurements. The \leff~ measurements from 20 to 46 \kevr~  suffer
from a low differential neutron-nuclear elastic scattering
cross-section compared to lower nuclear recoil energies
and thus have higher backgrounds.% because of scatters with the outside materials. 
Therefore, this results in larger statistical errors for measurements in this energy range. 
The detector was designed to 
minimize multiple scattering background 
for energies below 15 \kevr~ (see Figure \ref{fig:n_sim}). For 
10~\kevr, the \leff~ value agrees with previous experiments. For
energies below 10 \kevr, our results are lower than the Chepel 
\cite{Chepel:06} results and agree with the Aprile \cite{Aprile:08}  
values within errors. However, our results suggest a decreasing \leff~ 
with decreasing energy and not a constant as suggested by 
\cite{Sorensen:09}. Our results disagree with the \leff~ curves found
by comparing calibration data and Monte Carlo simulations (Figure \ref{fig:results_finalMC}) done by the XENON10 \cite{Sorensen:09} and 
ZEPLIN-III \cite{Lebedenko:08} collaborations. The ionization yield (Figure \ref{fig:QEQ0}) measured in this work is in agreement with  previous measurements \cite{Aprile:06} and the XENON10 Monte Carlo analysis \cite{Sorensen:09}.
% If instead of finding the \leff~ by the $\chi^2$ minimization method we simply compare the peak in the data  with the expected energy while taking into account the trigger and S1 finding efficiency, the \leff~ value  has no significant change for energies down to 10 \kevr, below this energy the \leff~ values shifts up to 35\% at 4 \kevr.

The theoretical \leff~ models by Lindhard \cite{Lindhard:63} and 
Hitachi \cite{Hitachi:05} fail to explain our measurements. The 
theoretical model presented in this paper, which includes the effect 
of escape electrons, fits our measurements within errors. 
This \leff~model is a simple one and several corrections can be done. 
As pointed out in \cite{Hitachi:07}, for LXe the Lindhard model is a crude approximation below 15 \kevr, 
and so the model could be improved below this energy. A more general expression for $q_{\textrm{esc}}$ could also be used that includes
different values of $\alpha$ for nuclear recoils and electron recoils  \cite{Dahl:09-phd} 
as well as any energy dependence of  $\alpha$.

\begin{figure}[htbp]
	\centering
		\includegraphics[height=3in]{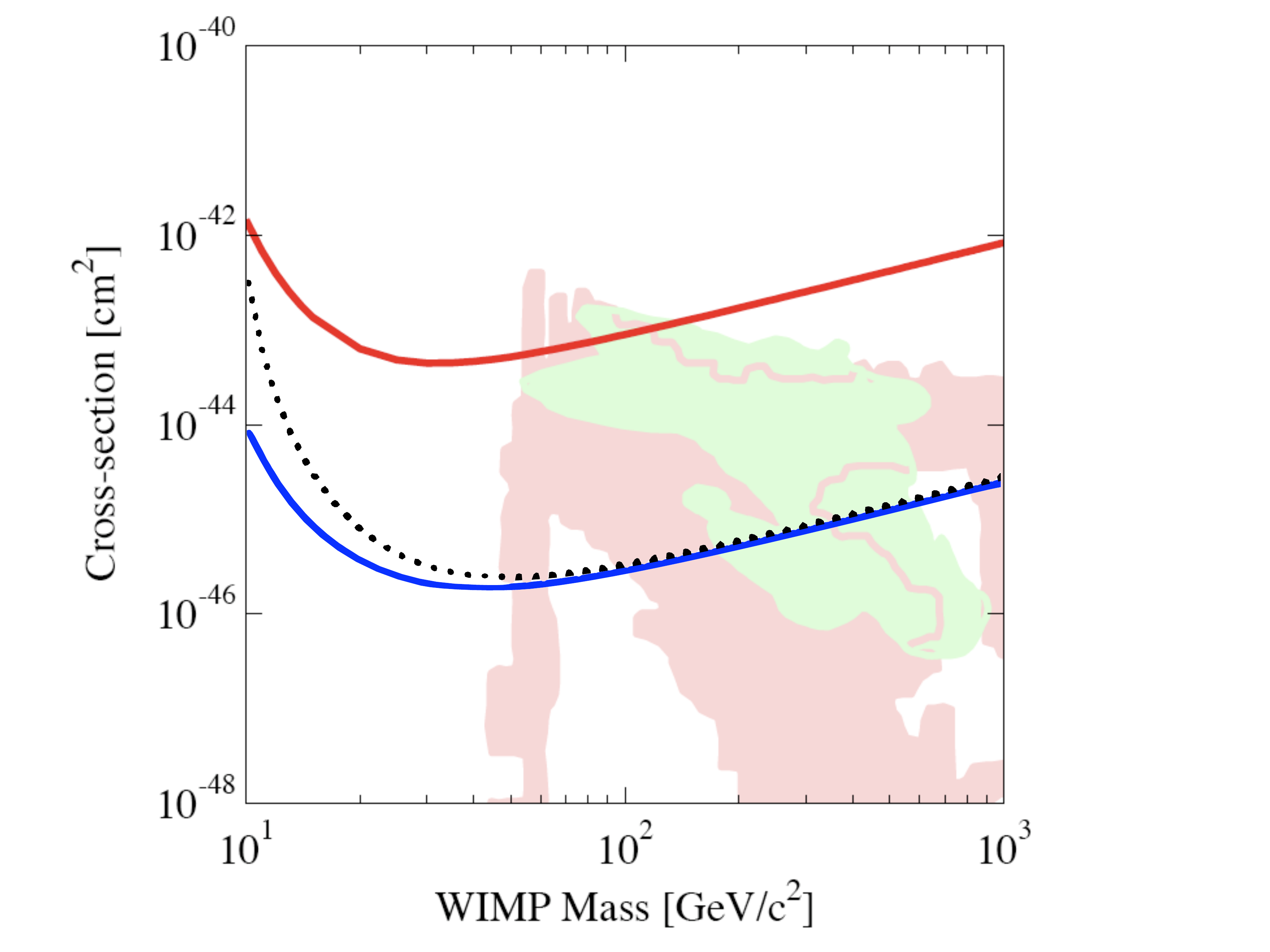}
	\caption{(Color online) Projected spin independent dark matter limits for a LXe detector with 30,000~kg~day exposure and 0.95 to 5.7 keVee energy window. The bottom solid line shows the limit with \leff=0.19 (5 to 30 \kevr~ window) while the dotted line shows the limit calculated with the measured \leff~(8.4 to 39.0 \kevr~ window).
 Also plotted are the  XENON10 result \cite{Angle:08} (top solid line) and the regions predicted by  \cite{Baltz:04:limit} (red shaded region) and \cite{Trotta:08} (green shaded region). Plot generated using \cite{dmplotter}.}
	\label{fig:XE_limit_paper}
\end{figure}

The \leff~ energy dependence measured in this work affects the dark matter limits set by LXe detectors. Figure \ref{fig:XE_limit_paper} shows the XENON10 result \cite{Angle:08} (top solid line)  found using \leff=0.19, as well as the projected spin-independent limits using \leff=0.19 (solid blue line) and the \leff~from this work (dotted line). The latter two curves are  90\% confidence limits based on Feldman-Cousins unified approach \cite{Feldman:98} for a LXe detector with a 30,000~kg~day exposure with no backgrounds in the 0.95 to 5.7 keVee energy window. This energy window corresponds to 5 to 30 \kevr~using \leff=0.19, and to 8.4 to 39.0 \kevr~using the measured \leff~ in this work. At a WIMP mass of 10 GeV/c$^2$, the WIMP nucleon cross-section limit  is more than an order of magnitude higher with the measured \leff, compared to the limit found assuming \leff=0.19. At 50 GeV/c$^2$ the WIMP nucleon cross-section limit is only 35\% higher and is 20\% higher above 100~GeV/c$^2$.

\section{Summary}
\label{sec:summary}
This work presents a new \leff~ measurement for energies between 4 
and 66 \kevr. This measurement was done using a single phase ($S1$ 
signal only) and a dual phase detector ($S1$ and $S2$ signals) to 
understand the detector's response at low energies. Each energy 
measured was repeated with at least three different electric fields 
finding no clear dependence of nuclear recoil scintillation  yield at low energies. We also present a 
theoretical \leff~model including nuclear quenching, bi-excitonic 
collisions and escape electrons that agrees with our results. 
Our \leff~ results and model suggest a decreasing \leff~ with 
decreasing energy. This result changes the  spin 
independent limit as shown in Figure \ref{fig:XE_limit_paper}. 
Although our \leff~result significantly changes the cross-section limit at low WIMP masses, the limit is only changed by 20\% for WIMP masses above 100 GeV/c$^2$.

In addition, we present the results for the nuclear recoil ionization yield 
and calculated the fraction of  escape electrons 
for nuclear recoil energies between 4 and 66 \kevr.

\begin{acknowledgments}
% \section{Acknowledgments}
	The authors would like to thank  George Andrews and Kevin 
Charbonneau for their assistance in running the neutron generator. 
The authors would also like to thank the Yale University Biomedical 
High Performance Computing Center where the simulations were 
performed. We thank Aaron Manalaysay for useful comments on the manuscript.
This work was supported by National Science Foundation grant  \#PHY-0800526.
\end{acknowledgments}

%\pagebreak
\bibliographystyle{h-physrev3} % I like
\bibliography{/Users/am542/References/thesis/research_ref}

\begin{thebibliography}{10}

\bibitem{Jungman:96}
G.~Jungman, M.~Kamionkowski, and K.~Griest,
\newblock Phys. Rep. {\bf 267}, 195 (1996).

\bibitem{Alner:07}
G.~Alner {\em et~al.},
\newblock Astropart. Phys. {\bf 28}, 287 (2007).

\bibitem{Angle:08}
J.~Angle {\em et~al.},
\newblock Phys. Rev. Lett. {\bf 100}, 021303 (2008).

\bibitem{Lebedenko:08}
V.~Lebedenko {\em et~al.},
\newblock arXiv {\bf 0812.1150} (2008).

\bibitem{Arneodo:00}
F.~Arneodo {\em et~al.},
\newblock Nucl. Inst. and Meth. A {\bf 449}, 147 (2000).

\bibitem{Akimov:02}
D.~Akimov {\em et~al.},
\newblock Phys. Lett. B {\bf 524} (2002).

\bibitem{Aprile:05}
E.~Aprile {\em et~al.},
\newblock Phys. Rev. D. {\bf 72} (2005).

\bibitem{Chepel:06}
V.~Chepel {\em et~al.},
\newblock Astropart. Phys. {\bf 26}, 58 (2006).

\bibitem{Aprile:08}
E.~Aprile {\em et~al.},
\newblock Phys. Rev. C {\bf 79}, 045807 (2009).

\bibitem{Aprile:06}
E.~Aprile {\em et~al.},
\newblock Phys. Rev. Lett. {\bf 97} (2006).

\bibitem{Sorensen:08}
P.~Sorensen,
\newblock {\em A position-sensitive liquid xenon time projection chamber for
  direct detection of dark matter: the Xenon10 experiment},
\newblock PhD thesis, Brown University, 2008.

\bibitem{Sorensen:09}
P.~Sorensen {\em et~al.},
\newblock Nucl. Inst. and Meth. A {\bf 601}, 339 (2009).

\bibitem{Chichester:07}
D.~Chichester, J.~Simpson, and M.~Lemchak,
\newblock J. Radioanal. Nucl. Chem. {\bf 271}, 629 (2007).

\bibitem{Lippincott:08}
W.~Lippincott {\em et~al.},
\newblock Phys. Rev. C {\bf 78}, 035801 (2008).

\bibitem{Nikkel:08}
J.~Nikkel, R.~Hasty, W.~Lippincott, and D.~McKinsey,
\newblock Astropart. Phys. {\bf 29}, 161 (2008).

\bibitem{Ni:07}
K.~Ni {\em et~al.},
\newblock Nucl. Inst. and Meth. A {\bf 582}, 569 (2007).

\bibitem{Geant4:03}
S.~Agostinelli and et~al,
\newblock Nucl. Inst. and Meth. B {\bf 506}, 250 (2003).

\bibitem{Pignil:07}
M.~Pignil, M.~Herman, P.~Oblozinsky, and D.~Rochman,
\newblock Brookhaven National Laboratory Report No. BNL-79261-2007-IR, 2007
  (unpublished).

\bibitem{Manzur:09}
A.~Manzur,
\newblock {\em Relative scintillation efficiency of liquid xenon in the XENON10
  direct dark matter search},
\newblock PhD thesis, Yale University, 2009.

\bibitem{PDG:06}
W.~M.~Yao {\em et~al.} [Particle Data Group],
\newblock J. Phys. G {\bf 33}, 1 (2006).

\bibitem{Lindhard:63}
J.~Linhard, M.~Scharff, and P.~Thomsen,
\newblock Mat. Fys. Medd. Dan. Vid. Selsk {\bf 33} (1963).

\bibitem{Doke:02}
T.~Doke, A.~Hitachi, J.~Kikuchi, and K.~Masuda,
\newblock Jpn. J. Appl. Phys. {\bf 41}, 1538 (2002).

\bibitem{Shutt:pri}
T.~Shutt,
\newblock Private communication  (2007).

\bibitem{Dahl:09-phd}
C.~Dahl,
\newblock {\em The physics of background discrimination in liquid xenon, and
  the first results from XENON10 in the hunt for WIMP dark matter},
\newblock PhD thesis, Princeton University, 2009.

\bibitem{Hitachi:05}
A.~Hitachi,
\newblock Astropart. Phys. {\bf 24}, 247 (2005).

\bibitem{Mei:08}
D.-M. Mei, Z.-B. Yin, L.~Stonehill, and A.~Hime,
\newblock Astropart. Phys. {\bf 30}, 12 (2008).

\bibitem{Birks:51}
J.~Birks and F.~Black,
\newblock Proc. Phys. Soc. A {\bf 64}, 511 (1951).

\bibitem{SRIM:09}
SRIM,
\newblock Particle interactions with matter, 2009.

\bibitem{Hitachi:07}
A.~Hitachi,
\newblock J. Phys: Conf. Ser. {\bf 65} (2007).

\bibitem{Feldman:98}
G.~Feldman and R.~Cousins,
\newblock Phys. Rev. D. {\bf 57}, 3873 (1998).

\bibitem{Baltz:04:limit}
E.~Baltz and P.~Gondolo,
\newblock JHEP {\bf 10}, 052 (2004).

\bibitem{Trotta:08}
R.~Trotta, F.~Feroz, M.~Hobson, L.~Roszkowski, and R.~Ruiz-de Austri,
\newblock JHEP {\bf 12}, 024 (2008).

\bibitem{dmplotter}
DMTOOLS,
\newblock http://dmtools.brown.edu, http://dmtools.berkeley.edu, 2009.

\end{thebibliography}

\end{document}